\documentclass[letterpaper,english,aps,prx,superscriptaddress,twocolumn,amsfonts, amssymb]{revtex4}
\usepackage[T1]{fontenc}
\pdfoutput=1
\usepackage{textcomp}
\usepackage{mathrsfs}
\usepackage{amsmath}
\usepackage{amssymb}
\usepackage{graphicx}
\usepackage{esint}
\usepackage{wasysym}

\makeatletter
\usepackage[english]{babel}

\usepackage[bookmarks=true,colorlinks,linkcolor=blue,urlcolor=blue,citecolor=blue]{hyperref}

\usepackage{epsfig,graphicx,psfrag,amsmath,amssymb,float}
\usepackage[caption=false]{subfig}

\@ifundefined{showcaptionsetup}{}{%
 \PassOptionsToPackage{caption=false}{subfig}}
\usepackage{subfig}
\makeatother

\usepackage{babel}

\begin{document}
\title{SU(4) Heisenberg model on the honeycomb lattice with exchange-frustrated
perturbations: Implications for twistronics and Mott insulators}
\author{W. M. H. Natori}
\affiliation{Instituto de F\'{i}sica de S\~{a}o Carlos, Universidade de S\~{a}o Paulo, S\~{a}o
Carlos, SP, 13560-970, CP 369, Brazil}
\affiliation{Blackett Laboratory, Imperial College London, London SW7 2AZ, United
Kingdom}
\author{R. Nutakki}
\affiliation{Department of Physics, University of Bath, Claverton Down, Bath, BA2
7AY, United Kingdom}
\author{R. G. Pereira}
\affiliation{International Institute of Physics and Departamento de F\'{i}sica Te\'{o}rica
e Experimental, Universidade Federal do Rio Grande do Norte, Natal-RN,
59078-970, Brazil}
\author{E. C. Andrade}
\affiliation{Instituto de F\'{i}sica de S\~{a}o Carlos, Universidade de S\~{a}o Paulo, S\~{a}o
Carlos, SP, 13560-970, CP 369, Brazil}
\begin{abstract}
The SU(4)-symmetric spin-orbital  model   on the honeycomb
lattice was recently studied in connection to correlated insulators
such as the $e_{g}$ Mott insulator Ba$_{3}$CuSb$_{2}$O$_{9}$ and
the insulating phase of magic-angle twisted bilayer graphene 
at quarter filling. Here we provide a unified discussion
of  these systems by investigating an extended model that includes the effects of Hund's coupling and anisotropic, orbital-dependent  exchange interactions. Using a combination of mean-field theory, linear flavor-wave theory, and variational Monte Carlo,
we show that this model harbors a quantum spin-orbital liquid
  over a wide parameter regime around the SU(4)-symmetric point.
For large Hund's coupling,  a ferromagnetic antiferro-orbital ordered state  appears, while
a valence-bond crystal combined with a vortex orbital state 
is stabilized by dominant orbital-dependent exchange interactions.
\end{abstract}
\maketitle

\section{Introduction\label{sec:Introduction}}

Kugel-Khomskii (KK) models   \citep{Kugel1982} are effective Hamiltonians  with couplings
between spin and orbital degrees of freedom that describe various phenomena
in transition metal oxides  \cite{Imada1998,Tokura2000}. 
Recently, the applications of KK models have been extended  to  Mott insulators with strong spin-orbit coupling
\citep{Khaliullin2005}, iron-pnictide superconductors \citep{Kruger2009},
Coulomb impurity lattices designed with scanning tunneling microscope
\citep{Dou2016}, and cold atom systems \citep{Ueda2018}. In realistic
KK models, the interplay between orbital configuration and lattice geometry
generally constrains the virtual electron transfers and generates
exchange frustration in the form of  bond-dependent and anisotropic spin-orbital
interactions \cite{Khaliullin2005}. This kind of exchange enhances quantum fluctuations
even in unfrustrated lattices \citep{Feiner1997}, leading to  the
expectation that   KK models may present exotic orders, valence
bond crystals (VBCs), or even quantum spin-orbital liquids (QSOLs)
as their ground states \cite{Nussinov2015}.

\textcolor{black}{The most well-studied examples of KK models display
two-orbital degeneracy and can be implemented in three distinct solid-state
platforms. Historically, the first one arises in Mott insulators
with $e_{g}$   orbitals \citep{Kugel1982}, where
the orbital Hilbert space is spanned by $d_{3z^{2}-r^{2}}$ and $d_{x^{2}-y^{2}}$
orbitals \citep{Reynaud2001,Mostovoy2002,Penc2003,Vernay2004,Reitsma2005}.
The second platform comprises $t_{2g}$ Mott insulators with 4/5$d^{1}$
magnetic species, in which the strong spin-orbit coupling (SOC) favors
a low-energy   $j=3/2$ multiplet
\citep{Chen2010}. These models can be alternatively
expressed in terms of pseudospins and pseudo-orbitals that mimic the
$e_{g}$ operators \citep{Natori2016,Romhanyi2017,Natori2017,Yamada2018,Natori2018}.
Lastly, two-orbital degenerate KK models were proposed as relevant
descriptions for correlated insulators observed in twistronic systems
\citep{Cao2018a,Cao2018b}. This proposal hangs upon the validity
of }Wannier orbitals to reproduce the twist-induced flat bands. If
this is the case and the interactions are sizable enough to describe
these systems in the strong-coupling regime, then KK Hamiltonians
naturally arise as minimal models for their insulating phases \citep{Xu2018,Venderbos2018,Yuan2018,Zhang2019,Classen2019,Schrade2019,Wu2019,ZhangB2019}.

One example of two-orbital KK model is the SU(4) Heisenberg model, which 
is receiving renewed interest due to suggested implementations in the three 
solid-state platforms described above \citep{Smerald2014,Kugel2015,Yamada2018,Natori2018,Venderbos2018,Yuan2018,Zhang2019,Classen2019,Schrade2019,Wu2019,ZhangB2019,Xu2018}.
Although the model is not exchange-frustrated, the higher symmetry fosters liquid ground states 
as first noted in SU(N) ``spin'' models   in the large-$N$ limit \citep{Affleck1988A,Arovas1988,Read1991}.
A specific study of the SU(4) Heisenberg model on the honeycomb lattice
was performed in Ref. \citep{Corboz2012} using several numerical
and analytical techniques. The combination of exact diagonalization
(ED) and Variational Monte Carlo (VMC) provided good evidence in
favor of a $\pi$-flux  QSOL  with fermionic excitations similar to the ones obtained
in large-$N$ theories. The experimental motivation of \citep{Corboz2012}
was the $e_{g}$ system Ba$_{3}$CuSb$_{2}$O$_{9}$, in which Cu$^{2+}$
ions were proposed to form layered honeycomb lattices \citep{Nakatsuji2012}.
Other theoretical descriptions of the same compound also regarded
the SU(4) Heisenberg model as relevant, but included exchange-frustrated
terms induced by orbital-dependent virtual hopping processes \citep{Nasu2013,Smerald2014}. In the last year, two new platforms
for the SU(4) Heisenberg model were proposed: the analogues of Kitaev
materials with 4/5$d^{1}$ magnetic species (e.g., $\alpha$-ZrCl$_{3}$)
\citep{Yamada2018,Natori2018} and the Mott phase of twisted bilayer
graphene (TBG) \citep{Venderbos2018}.

\begin{figure}
\begin{centering}
\includegraphics[width=1\columnwidth]{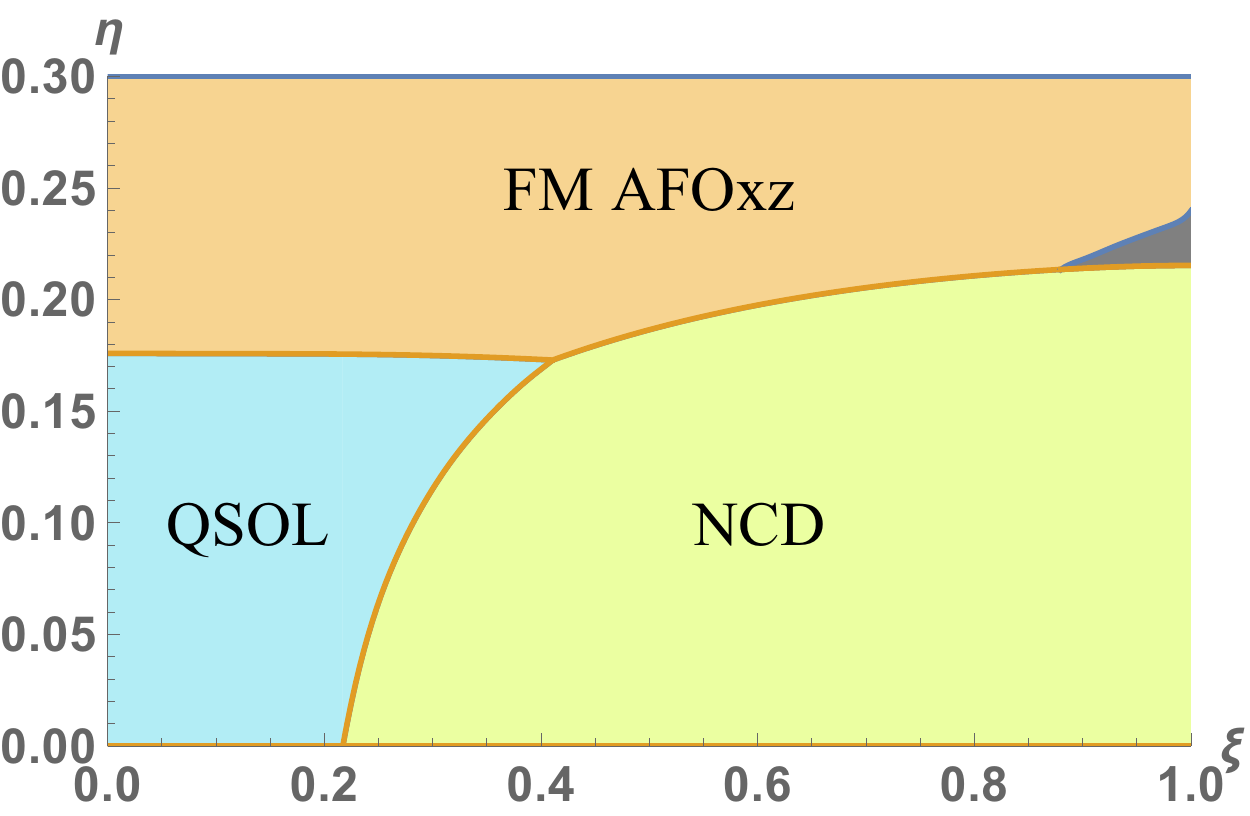}
\par\end{centering}
\caption{\label{fig:phase_diagram} Phase diagram of the Kugel-Khomskii model as a 
function of the ratio of hopping parameters $\xi=t'/t$ and the dimensionless Hund's coupling 
parameter $\eta=J_H/U$. There are three distinct phases: a quantum
spin-orbital liquid (QSOL, in blue), a noncollinear ordering of spin dimers (NCD, in green) 
and a ferromagnetic state with staggered orbital order (FM AFOxz, in orange).
A region of instability of the last phase is indicated in gray and discussed in Subsection
\ref{subsec:LFWT}.}
\end{figure}

The purpose of this paper is to study  the effects of exchange-frustrated and
Hund's coupling induced interactions on the SU(4) Heisenberg model
on the honeycomb lattice. We present a detailed analysis of a KK model 
derived independently in Refs. \citep{Venderbos2018} and \citep{Smerald2014} using mean-field theory (MFT),
linear flavor-wave theory (LFWT) (complemented by a variational study considering the Huse-Elser 
wavefunction \citep{Huse1988,Ferrari2017}) and VMC. Our main results are  summarized in Fig. \ref{fig:phase_diagram}. 
Our study  corroborates  the existence of a stable QSOL phase around the SU(4)-symmetric point 
studied in Ref. \citep{Corboz2012}. For larger values of the SU(4)-symmetry-breaking interactions, we find either 
a two-sublattice  state with ferromagnetic order for the spin degrees of freedom or a VBC phase of spin dimers coupled
to a three-sublattice   vortex orbital state. Our phase diagram agrees qualitatively 
with the one obtained in \cite{Smerald2014} by exact diagonalization of the same model on small clusters.

The remaining sections are organized as follows. We present the local degrees of freedom and the
KK model in Sec. \ref{sec:Microscopical-model}. Besides fixing the notation, this section also discusses
the properties of the orbital degrees of freedom and symmetries of the model that will be 
relevant for our subsequent analysis. Sec. \ref{sec:Ordered-States} identifies possible ordered
ground states of the KK model using MFT, expanding the phase diagram presented in Ref. \citep{Venderbos2018}. The effects of quantum 
fluctuations on these states are then evaluated within LFWT. The QSOL proposed in Ref. \citep{Corboz2012} 
and possible VBCs ground states of this KK model are studied within VMC 
as presented in Sec. \ref{sec:QSOL_VBC}. The phase diagram in Fig. \ref{fig:phase_diagram}
is constructed through the combination of the LFWT and VMC energetics studies. 
The relevance of our results and perspectives for future work are provided in Sec. \ref{sec:Discussion}.

\section{Microscopic Models \label{sec:Microscopical-model}}

\subsection{Local degrees of freedom \label{subsec:degrees-of-freedom}}

Let us start with a brief description of the local degrees of freedom
of the magnetic species that we are investigating. We assign to each site a 
spin 1/2 as well as an orbital degree of freedom corresponding to quantum numbers $S^z=\pm\frac12$ and $\tau^{z}=\pm \frac{1}{2}$, respectively.  The Hilbert space of each site $i$
is then spanned by four states $\left|S_{i}^{z},\tau_{i}^{z}\right\rangle $
(often called \emph{colors}) which are labeled as
\begin{align}
\left|1\right\rangle =\left|\frac{1}{2},\frac{1}{2}\right\rangle , & \quad \left|2\right\rangle =\left|-\frac{1}{2},\frac{1}{2}\right\rangle ,\nonumber \\
\left|3\right\rangle =\left|\frac{1}{2},-\frac{1}{2}\right\rangle , &\quad \left|4\right\rangle =\left|-\frac{1}{2},-\frac{1}{2}\right\rangle .\label{eq:colors}
\end{align}
The operators for spin ($\mathbf{S}_{i}$) and orbital ($\boldsymbol{\tau}_{i}$)
obey the usual SU(2) algebra $\left[S_{i}^{\alpha},S_{j}^{\beta}\right]=i\epsilon^{\alpha\beta\gamma}S_{i}^{\gamma}\delta_{ij}$,
$\left[\tau_{i}^{\alpha},\tau_{j}^{\beta}\right]=i\epsilon^{\alpha\beta\gamma}\tau_{i}^{\gamma}\delta_{ij}$
and $\left[S_{i}^{\alpha},\tau_{j}^{\beta}\right]=0$ and are represented by the Pauli matrices in their respective spaces.

The orbital   degree of freedom may describe, for instance,  a low-energy  
$e_{g}$ doublet $\left\{ d_{3z^{2}-r^2},d_{x^{2}-y^{2}}\right\} $ in Mott insulators 
with octahedral crystal field \cite{Khaliullin2005}.  Alternatively, it may refer to 
$p_x$ and $p_y$ orbitals  in optical lattices \cite{Wu2008}  or twistronic systems  \cite{Venderbos2018}. 
The doublets in these two cases constitute the orthogonal eigenstates of the $\tau^z$ operator. 
We will be concerned only with Hamiltonians that remain invariant under $C_{3}$ rotations around 
the normal axis of the honeycomb lattice. Similar $C_{3}$ rotations on the internal orbital subspace spanned 
by $(\tau^{z},\tau^{x})$ accounts for the effect of the spatial transformations on the orbitals.
The remaining component $\tau^{y}$ has a distinct role that is more easily seen by the
effect of the time-reversal operator $\Theta$. In the orbital space, $\Theta$ reduces
to a complex conjugation and leads to $\Theta\boldsymbol{\tau}\Theta^{-1}=\left(\tau^{x},-\tau^{y},\tau^{z}\right)$. 
The physical interpretation of $\tau^{y}$ is that it distinguishes between states with 
different orbital chiralities \citep{Yuan2018} and is, therefore, related to
orbital-magnetic orders \cite{Venderbos2018}.

\begin{figure}
\begin{centering}
\includegraphics[width=1\columnwidth]{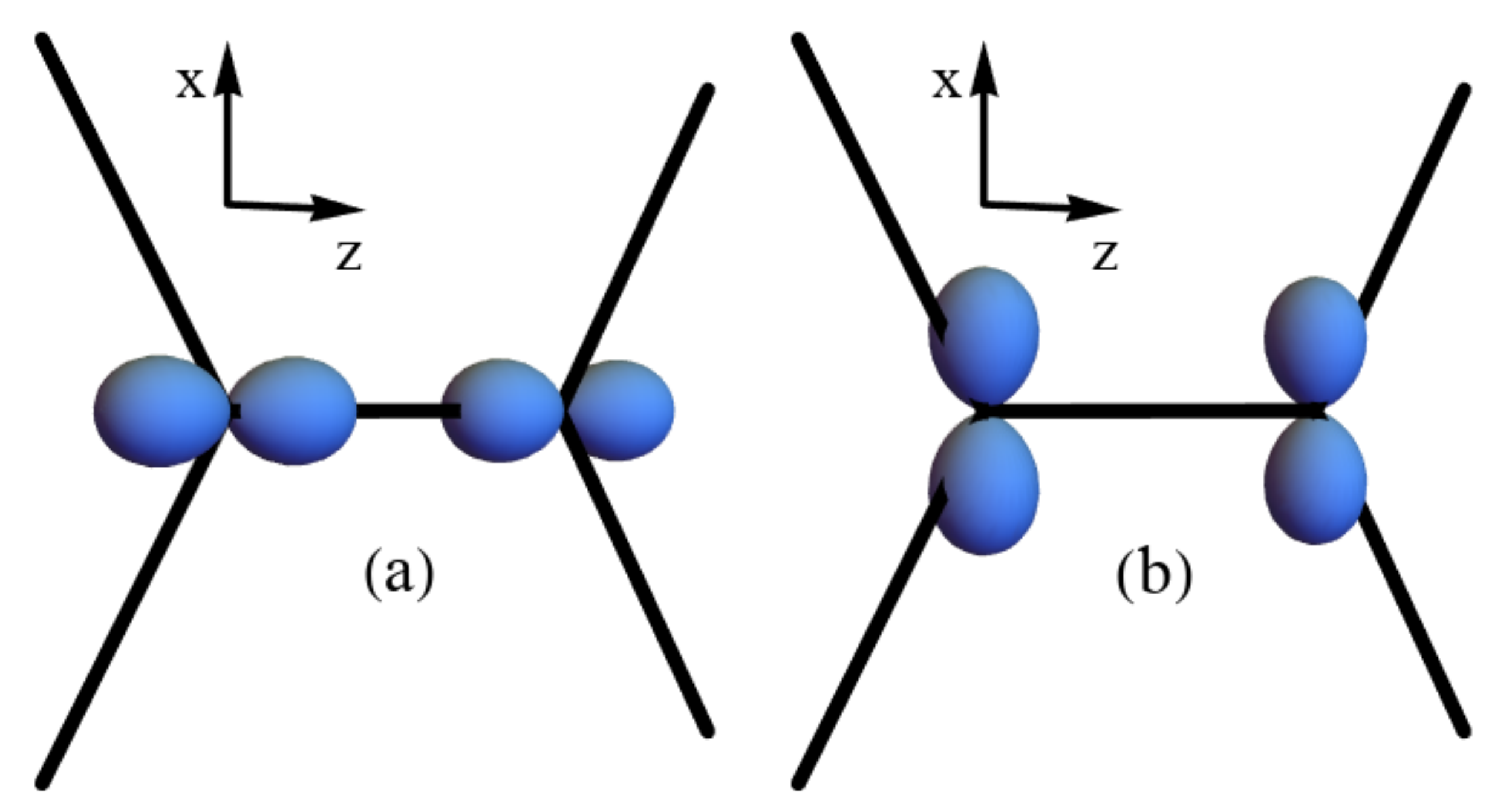}
\par\end{centering}
\caption{\label{fig:p_eg_fig} Representation of (a) $p_{z}$ and (b) $p_{x}$ 
orbitals in the $zx$ plane. The orbitals in Fig. (a) 
overlap with each other analogously to a $\sigma$-bond in organic chemistry. 
Analogously, Fig. (b) depicts active orbitals similar to a $\pi$-bond.}

\end{figure}

\subsection{Kugel-Khomskii Model   \label{subsec:derivation}}

We now   introduce the minimal model proposed for the Mott insulating
phase of Ba$_{3}$CuSb$_{2}$O$_{9}$ and TBG in Refs. \citep{Smerald2014,Venderbos2018}. Despite the different nature of the orbitals in these systems, their in-plane symmetries enable one to assign the same Hubbard model in both cases. The
interactions $H_{I}$ are restricted to be onsite:
\begin{align}
H_{I} & =U\sum_{i}\sum_{\alpha=1,2}n_{i,\alpha,\uparrow}n_{i,\alpha,\downarrow}+\left(U-2J_{H}\right)\sum_{i}n_{i,1}n_{i,2}\nonumber \\
 &\quad +J_{H}\sum_{i,s,s^{\prime}}c_{i,1,s}^{\dagger}c_{i,2,s^{\prime}}^{\dagger}c^{\phantom\dagger}_{i,1,s^{\prime}}c^{\phantom\dagger}_{i,2,s}\nonumber \\
 & \quad+J_{H}\sum_{i,\alpha\neq\beta}c_{i,\alpha,\uparrow}^{\dagger}c_{i,\alpha,\downarrow}^{\dagger}c^{\phantom\dagger}_{i,\beta,\downarrow}c^{\phantom\dagger}_{i,\beta,\uparrow},\label{eq:HI}
\end{align}
in which $c_{i,\alpha,s}=c_{\alpha,s}(\mathbf r_i)$ is the annihilation operator of an electron
at position  $\mathbf r_i$ on the honeycomb lattice with orbital state   $\alpha=1,2$ (corresponding to $\tau^z=\pm\frac12$, respectively) and   spin $s=\uparrow,\downarrow$ (for $S^z=\pm\frac12$).
We also introduce the number operator for a given orbital as $n_{i,\alpha}=\sum_{s}c_{i,\alpha,s}^{\dagger}c_{i,\alpha,s}$
and two parameters for electrostatic interactions: the direct Coulomb
repulsion $U>0$ and Hund's coupling $J_{H}>0$. The tunneling between
nearest neighbors on the honeycomb lattice is modeled by the tight-binding
Hamiltonian
\begin{equation}
H_{\text{TB}}=\sum_{i\in\textrm{A}}\sum_{\alpha,\beta}\sum_{\gamma=1}^{3}\sum_{s}c_{\alpha,s}^{\dagger}(\mathbf{r}_{i})\hat{h}_{\alpha\beta}^{(\gamma)}c_{\beta,s}\left(\mathbf{r}_{i}+\frac{\hat{\mathbf{e}}_{\gamma}}{\sqrt{3}}\right)+\text{h.c.},\label{eq:kinetic}
\end{equation}
where $\hat{\mathbf{e}}_{1}=\hat{\mathbf{z}},\,\hat{\mathbf{e}}_{2}=-\frac{1}{2}\hat{\mathbf{z}}+\frac{\sqrt{3}}{2}\hat{\mathbf{x}},\,\hat{\mathbf{e}}_{3}=-\frac{1}{2}\hat{\mathbf{z}}-\frac{\sqrt{3}}{2}\hat{\mathbf{x}}$ are unit vectors in the $zx$ plane, 
and $i$ runs over the  A sublattice, i.e., the triangular Bravais lattice. Here we have set the lattice spacing of the honeycomb lattice to $1$. The matrix
$\hat{h}_{\alpha\beta}^{(\gamma)}$ depends on the overlap between the orbitals at positions $\mathbf r_i$ and $\mathbf r_j$, connected by a link in the direction  $\mathbf r_{ij}=\mathbf r_j-\mathbf r_i=\frac1{\sqrt3}\hat{\mathbf e}_\gamma$. 
Using an analogy with
organic chemistry, we can think of two types of hoppings involving
$p$ orbitals that are connected by $\sigma$ or $\pi$ bonds,
as indicated in Fig. \ref{fig:p_eg_fig}. The matrix $\hat{h}_{\alpha\beta}^{(\gamma)}$
is then parameterized as \citep{Smerald2014,Venderbos2018}
\begin{align}
\hat{h}_{\alpha\beta}^{(\gamma)} & =t+2t^{\prime}\hat{\mathbf{e}}_{\gamma}\cdot\boldsymbol{\tau}\nonumber \\
 & =\left[t_{\sigma}\left(\frac{1}{2}+\hat{\mathbf{e}}_{\gamma}\cdot\boldsymbol{\tau}\right)+t_{\pi}\left(\frac{1}{2}-\hat{\mathbf{e}}_{\gamma}\cdot\boldsymbol{\tau}\right)\right],\label{eq:T}
\end{align}
where 
 $t_{\sigma,\pi}\equiv t\pm t^{\prime}$.

\begin{widetext}
The KK model  derived  within  second-order perturbation theory in the regime $t,t'\ll U,J_H$ reads  \begin{align}
\mathcal{H}\left(\xi,\eta\right) & =\sum_{\langle ij\rangle}\left\{ J_{1}\mathcal{P}_{ij}^{1}\left[(1-\xi^2)\mathcal{Q}_{ij}^{+}-2\left(1+\xi^{2}\right)\left(\mathcal{P}_{ij}^{+-}+\mathcal{P}_{ij}^{-+}\right)\right] \right.\nonumber \\
 & \qquad-J_{2}\mathcal{P}_{ij}^{0}\left[8(1-\xi^2)\tau_{i}^{y}\tau_{j}^{y}+2(1+\xi)^{2}\mathcal{P}_{ij}^{++}+2(1-\xi)^{2}\mathcal{P}_{ij}^{--}+2\left(1+\xi^2\right)\left(\mathcal{P}_{ij}^{+-}+\mathcal{P}_{ij}^{-+}\right)\right] \nonumber\\
 & \left.\qquad-J_{3} \mathcal{P}_{ij}^{0}\left((1-\xi^2)\mathcal{Q}_{ij}^{-}+2(1+\xi)^{2}\mathcal{P}_{ij}^{++}+2(1-\xi)^{2}\mathcal{P}_{ij}^{--}\right) \right\}
 .\label{eq:complete_Hamiltonian}
\end{align}
\end{widetext}
Here we defined the dimensionless parameters $\xi=t'/t$ and $\eta=J_{H}/U$ and the exchange coupling
constants $
J_{1}=\frac{J}{1-3\eta}$, $J_{2}=\frac{J}{1-\eta}$,   $  J_{3}=\frac{J}{1+\eta}
$, where $J=t^2/U$. 
As usual in KK models, the spin part of the  interaction between the electrons at sites $i$ and $j$ is written in terms of the projectors onto states with total spin $S=0$ and $S=1$: 
\begin{align}
\mathcal{P}_{ij}^{0} & =\frac{1}{4}-\mathbf{S}_{i}\cdot\mathbf{S}_{j},\qquad  
\mathcal{P}_{ij}^{1} =\mathbf{S}_{i}\cdot\mathbf{S}_{j}+\frac{3}{4}.
\end{align}
Notice that the  Hamiltonian   is invariant
under global spin SU(2) rotations.
In contrast, the orbital part of the interaction is in general anisotropic and bond dependent, as it involves the operators
\begin{align}
\mathcal{P}_{ij}^{\mu\nu} & \equiv\left(\frac{1}{2}+\mu\hat{\mathbf{e}}_{ij}\cdot\boldsymbol{\tau}_{i}\right)\left(\frac{1}{2}+\nu\hat{\mathbf{e}}_{ij}\cdot\boldsymbol{\tau}_{j}\right),\nonumber \\
\mathcal{Q}_{ij}^{+} & =4\left[\boldsymbol{\tau}_{i}\cdot\boldsymbol{\tau}_{j}-\left(\hat{\mathbf{e}}_{ij}\cdot\boldsymbol{\tau}_{i}\right)\left(\hat{\mathbf{e}}_{ij}\cdot\boldsymbol{\tau}_{j}\right)\right],\nonumber \\
\mathcal{Q}_{ij}^{-} & =4\left[\boldsymbol{\tau}_{i}\cdot\boldsymbol{\tau}_{j}-2\tau_{i}^{y}\tau_{j}^{y}-\left(\hat{\mathbf{e}}_{ij}\cdot\boldsymbol{\tau}_{i}\right)\left(\hat{\mathbf{e}}_{ij}\cdot\boldsymbol{\tau}_{j}\right)\right],\label{eq:orbital_operators}
\end{align}
where $\mu,\nu\in\{+,-\}$ and $\hat{\mathbf e}_{ij}=\hat{\mathbf e}_\gamma$ for  $\mathbf r_{ij}\parallel\hat{\mathbf e}_\gamma$. Explicitly, we can write the Hamiltonian as \begin{widetext}
\begin{align}
\mathcal{H} (\xi,\eta)& =\sum_{\langle ij\rangle}\left\{ \left(J_{1}+J_{2}\right)\left(1-\xi^2\right)\left(2\mathbf{S}_{i}\cdot\mathbf{S}_{j}+\frac{1}{2}\right)\left(2\boldsymbol{\tau}_{i}\cdot\boldsymbol{\tau}_{j}+\frac{1}{2}\right)\right.+\left(J_{1}-J_{3}\right)\left(1-\xi^2\right)\left(2\boldsymbol{\tau}_{i}\cdot\boldsymbol{\tau}_{j}+\frac{1}{2}\right)\nonumber \\
 & \qquad+2\left(J_{2}-J_{3}\right)\left(1-\xi^2\right)\left(2\mathbf{S}_{i}\cdot\mathbf{S}_{j}-\frac{1}{2}\right)\left(2\tau_{i}^{y}\tau_{j}^{y}+\frac{1}{2}\right)-\left(J_{1}-2\xi^2J_{2}-J_{3}\right)\left(2\mathbf{S}_{i}\cdot\mathbf{S}_{j}+\frac{1}{2}\right)\nonumber \\
 & \qquad+\xi\left(J_{2}+J_{3}\right)4\mathbf{S}_{i}\cdot\mathbf{S}_{j}\left(\hat{\mathbf{e}}_{ij}\cdot\boldsymbol{\tau}_{i}+\hat{\mathbf{e}}_{ij}\cdot\boldsymbol{\tau}_{j}\right)+\xi^2\left(J_{1}+J_{3}\right)8\mathbf{S}_{i}\cdot\mathbf{S}_{j}\left(\hat{\mathbf{e}}_{ij}\cdot\boldsymbol{\tau}_{i}\right)\left(\hat{\mathbf{e}}_{ij}\cdot\boldsymbol{\tau}_{j}\right)\nonumber\\
 & \qquad\left.+2\xi^2\left(3J_{1}-J_{3}\right)\left(\hat{\mathbf{e}}_{ij}\cdot\boldsymbol{\tau}_{i}\right)\left(\hat{\mathbf{e}}_{ij}\cdot\boldsymbol{\tau}_{j}\right)  -\left(J_{1}+2\xi^2J_{2}+J_{3}\right)\right\} .\label{eq:complete_Hamiltonian-1}
\end{align}
\end{widetext}

Let us first consider the model  with  $\xi=\eta=0$. In this case, the original two-orbital Hubbard model  in Eqs. (\ref{eq:HI}) and (\ref{eq:kinetic}) is invariant under  global  SU(4) 
color transformations. As a result,  at this point the KK model reduces to $\mathcal{H}(0,0)=\mathcal{H}_{\text{SU(4)}}-3NJ$, where  $N$ is the number of sites of the honeycomb lattice and $\mathcal{H}_{\text{SU(4)}}$ is the SU(4) Heisenberg model  given by
\begin{align}
 \mathcal{H}_{\text{SU(4)}}& =2J\underset{\langle ij\rangle}{\sum}\left(2\mathbf{S}_{i}\cdot\mathbf{S}_{j}+\frac{1}{2}\right)\left(2\boldsymbol{\tau}_{i}\cdot\boldsymbol{\tau}_{j}+\frac{1}{2}\right),\nonumber \\
 & =2J\underset{\langle ij\rangle}{\sum}\sum_{a,b=1}^{4}S_{a}^{b}(i)S_{b}^{a}(j).\label{eq:SU4} 
\end{align}
Here we introduce the color exchange operators
\begin{equation}
S_{a}^{b}(i)=\left|a\right\rangle _{i}\left\langle b\right|_{i},\label{eq:Sab}
\end{equation}
which can be recognized as the SU(4) generators \cite{Corboz2012}.  Any local spin-orbital operator that appears in Eq. \eqref{eq:complete_Hamiltonian} can be written as a linear combination of $S_{a}^{b}(i)$ operators. In particular, $\mathcal{H}_{\text{SU(4)}}$
is  proportional to the sum of color permutation operators $P_{ij}\equiv\sum_{a,b}S_{a}^{b}(i)S_{b}^{a}(j)$
over all nearest-neighbor bonds.

In the case of isotropic hopping ($\xi=0$) but nonzero Hund's coupling ($\eta>0$), the Hamiltonian is given by 
\begin{align}
\mathcal{H}\left(0,\eta\right) & =\sum_{\langle ij\rangle}\left[4J_{1}\mathcal{P}_{ij}^{1}\left(\boldsymbol{\tau}_{i}\cdot\boldsymbol{\tau}_{j}-\frac{1}{4}\right) \right.\nonumber\\
&\quad-8J_{2}\mathcal{P}_{ij}^{0}\left(\tau_{i}^{y}\tau_{j}^{y}+\frac{1}{4}\right)\nonumber \\
 & \left.\quad-4J_{3}\mathcal{P}_{ij}^{0}\left(\boldsymbol{\tau}_{i}\cdot\boldsymbol{\tau}_{j}-2\tau_{i}^{y}\tau_{j}^{y}+\frac{1}{4}\right)\right].
\end{align}
Thus, along the  $\xi=0$ line in parameter space, the model  retains an SU(2)$\times$U(1) symmetry. 
The U(1) symmetry is due to the conservation of the orbital chirality, as $\sum_i\tau^y_i$ commutes with the Hamiltonian.
 
For general values of $\xi$ and $\eta$, model \eqref{eq:complete_Hamiltonian} exhibits a 
global SU(2)$\times\mathbb Z_3$ symmetry, where the $\mathbb Z_3$ symmetry is associated with  
$\pm 120^\circ$ orbital rotations about $\tau^y$ accompanied by  the  rotation of the bond directions.  
The   bond-dependent hopping  $t^{\prime}$
introduces the  exchange-frustrated perturbations  given  in Eq. (\ref{eq:complete_Hamiltonian-1}). 
 Of particular interest are the points $\xi=\pm1$ with $\eta=0$, where Eq.
(\ref{eq:complete_Hamiltonian}) becomes
\begin{equation}
\mathcal{H}\left(\pm 1,0\right)=8J\sum_{\langle ij\rangle}\left(2\mathbf{S}_{i}\cdot\mathbf{S}_{j}+\frac{1}{2}\right)\mathcal{P}_{ij}^{++(--)}-6NJ.\label{eq:SU2Z3}
\end{equation}
The orbital interactions take this form  because Eq.
(\ref{eq:T}) involves a projector to either  $\sigma$ or $\pi$
bonds (Fig. \ref{fig:p_eg_fig}). Hence, the electrons
  interact with each other only if they   both occupy the 
orbital state which is an eigenstate of $\hat{\mathbf e}_{ij}\cdot\boldsymbol \tau$ 
with eigenvalue $\pm1/2$ for $t'=\pm t$, respectively.  This type of orbital 
dependence appears in compass models for $e_g$ orbitals \cite{Nussinov2015} or for 
$j=3/2$ states after projection of $t_{2g}$ states in the limit of strong 
spin-orbit coupling \cite{Natori2016,Romhanyi2017}.

\section{Ordered States\label{sec:Ordered-States}}

The first step to gain intuition of the phase diagram of the model in Eq. (\ref{eq:complete_Hamiltonian})
is to study ordered states with   MFT. In
this section, we   study classical ordered states that
are equivalent to a product state
\begin{equation}
\left|\Psi\right\rangle =\prod_{i}\left|\psi_{i}\right\rangle ,
\end{equation}
in which $\left|\psi_{i}\right\rangle $ is a linear combination of
the states in Eq. (\ref{eq:colors}). 
The method provides the phase diagram
in Fig. \ref{fig:Classical_diagram}(a), which extends the result of Ref. \cite{Venderbos2018} by including nonzero orbital-dependent hopping $t'$.    We shall then analyze the stability of the ordered states against quantum fluctuations using LFWT \cite{Joshi1999}.

\subsection{Mean Field Theory \label{subsec:Mean-Field}}

\begin{figure}
\begin{centering}
\subfloat{\centering{}\includegraphics[width=0.75\columnwidth]{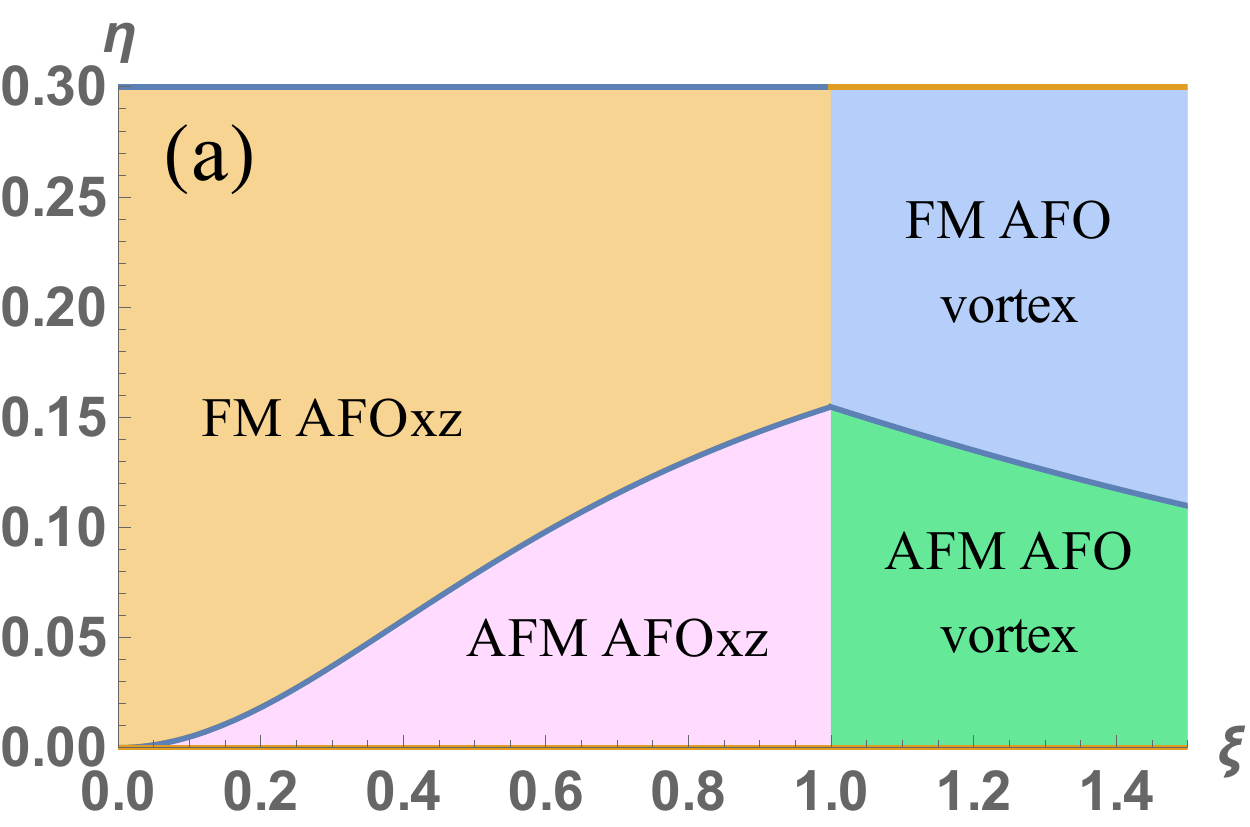}}
\par\end{centering}
\begin{centering}
\subfloat{\includegraphics[width=0.35\columnwidth]{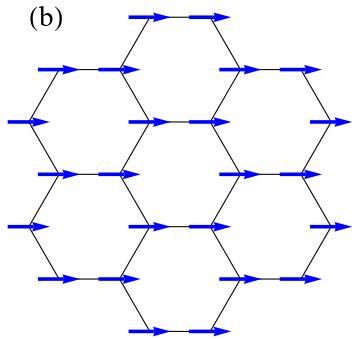}}\quad\subfloat{\includegraphics[width=0.35\columnwidth]{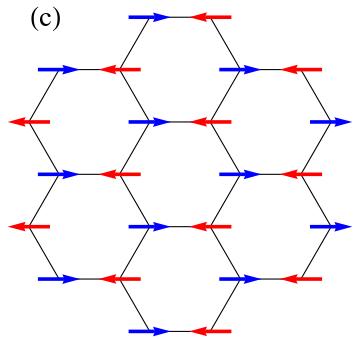}}
\par\end{centering}
\begin{centering}
\subfloat{\includegraphics[width=0.38\columnwidth]{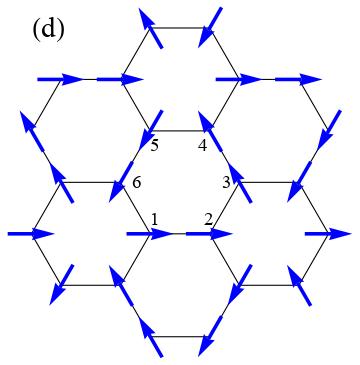}}\quad\subfloat{\includegraphics[width=0.38\columnwidth]{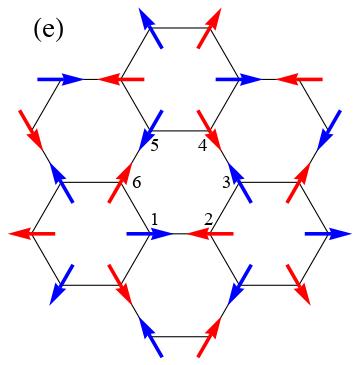}}
\par\end{centering}
\caption{\label{fig:Classical_diagram}Classical states in  the spin-orbital model Eq. (\ref{eq:complete_Hamiltonian}). (a) Mean-field phase
diagram. The spin sector may display ferromagnetic (FM) or antiferromagnetic (AFM) order. In (b) through (d), the arrows represent the orbital order within the plane $\left(\tau^{z},\tau^{x}\right)$: (b) ferro-orbital
(FOxz), (c) N\'{e}el (AFOxz), (d) ferro-orbital vortex and (e) antiferro-orbital
vortex. The numbers in (d) and (e) indicate the six-sublattice magnetic unit cell. }
\end{figure}

Our choice of ordered states is guided by the
symmetries discussed in Subsection \ref{subsec:derivation}. The SU(2)
symmetry and the absence of geometric frustration   suggest that, classically, the spins
form either  a ferromagnetic (FM) or an antiferromagnetic (AFM) order. 
Equation (\ref{eq:orbital_operators}) suggests that the orbitals 
may either align with $\tau^{y}$ or be contained in the $\left(\tau^{z},\tau^{x}\right)$
plane. By computing the classical energy for different orbital configurations, we find that only in-plane orbital-ordered 
states are competitive.  When  bond-independent  interactions   
dominate, the system   develops  ferro-orbital
(FOxz) or antiferro-orbital (AFOxz) in the $xz$ plane as illustrated  in Figs. \ref{fig:Classical_diagram}(b)
and (c). On the other hand, interactions proportional to $\left(\hat{\mathbf{e}}_{ij}\cdot\boldsymbol{\tau}_{i}\right)\left(\hat{\mathbf{e}}_{ij}\cdot\boldsymbol{\tau}_{j}\right)$
 favor the ``vortex''
orbital orders  displayed
in Figs. \ref{fig:Classical_diagram}(d) and (e). In fact, the orbital
vortex states are the best trial ground states of the compass model
on the honeycomb lattice \citep{Wu2008}. They also appear as the exact ground
state for a special point of the $JK\Gamma$ model  for  the honeycomb   iridates \citep{Chaloupka2015}.

We find that the mean field phase diagram is symmetric under $\xi\mapsto -\xi$. Figure   \ref{fig:Classical_diagram}(a) shows  the phase  diagram for $\xi>0$ and $0<\eta<0.3$. In the physically more relevant regime $\xi<1$, we  observe a competition between two states with AFOxz order distinguished by the FM or AFM spin order. In agreement with \cite{Venderbos2018}, the FM state has lower energy for $\xi\to0$ at any fixed $\eta$. The transition from AFOxz to the orbital vortex states across the line $\xi=1$  can be attributed to a six-sublattice orbital rotation symmetry of the model  discussed in \citep{Smerald2014}, which maps $\mathcal H(\xi,\eta)\mapsto \mathcal H(1/\xi,\eta)$ and  connects the collinear orbital phases
in Figs. \ref{fig:Classical_diagram}(b) and (c) to their vortex counterparts
in Fig. \ref{fig:Classical_diagram}(d) and (e) \citep{Chaloupka2015}.

\subsection{Linear Flavor Wave Theory \label{subsec:LFWT}}

A careful analysis of MFT indicates that the classical phase diagram
in Fig. \ref{fig:Classical_diagram}(a) is incorrect near the point $\xi=\eta=0$, 
corresponding to $\mathcal{H}_{\text{SU(4)}}$. The reason is that the expectation value  
$\left\langle \Psi\right.\left|\mathcal{H}_{\text{SU(4)}}\right|\left.\Psi\right\rangle $
is the same for any state in which $\left\langle \psi_{i}|\psi_{j}\right\rangle =0$,
i.e., whenever neighboring sites have different colors \citep{Corboz2012}.
The number of states satisfying this constraint increases exponentially
with the system size and flags the onset of a disordered
state. On the other hand, finite Hund's coupling is expected to favor 
spin ferromagnetism, in consistency with the FM AFOxz phase. 
It is then desirable to study the effect of quantum fluctuations on 
the energy and stability of this spin-orbital ordered state.

LFWT can be viewed as the analog of spin wave theory for spin-orbital models \citep{Joshi1999}. It allows estimates of the excitation dispersion, correction to the zero-point energy and reduction of the order parameter by quantum fluctuations ($\Delta M$) in a single formalism. It also provides some criteria for the stability of a given ordered phase. For example, the application of LFWT to AFM AFOxz leads to dispersion relations with complex frequencies at any $\eta\neq0$ and $\xi\neq0$. Such complex dispersion clearly indicates that the AFM AFOxz state is unstable and explains its absence in Fig. \ref{fig:phase_diagram}.  

We then study the FM AFOxz state, which is the ordered state with fixed colors $m_A=3$ and $m_B=1$ on the A and B sublattices, respectively. The Holstein-Primakoff transformation introduces three bosonic species per sublattice labeled by $b_{irm}$, in which $i$ indexes the unit cells, $r$ the sublattices and $m \neq m_r$ correspond to the colors in Eq. (\ref{eq:colors}). After replacing spin-orbital operators by their bosonic representations and truncating the Hamiltonian at the level of quadratic terms, the LFWT Hamiltonian is written in the Fourier space as
\begin{align}
\mathcal{H}_{\text{LFWT}} & =-2NJ_{1}\left(3+\xi^2\right)+\sum_{\mathbf{k}}B_{\mathbf{k}}^{\dagger}\mathcal{H}_{\mathbf{k}}B^{\phantom\dagger}_{\mathbf{k}}\nonumber \\
& \,\,\,\,\,\,-\frac{3}{2}N\left[\left(3+\frac{\xi^2}{2}\right)J_{1}-2\left(1+\xi^2\right)J_{2}\right. \nonumber\\
& \,\,\,\,\,\,\left.-\left(1+\xi^2\right)J_{3}\right],\label{eq:HLFWT}
\end{align}
where $\mathbf{k}$ lies in the Brillouin zone. Here $B_{\mathbf{k}}^{\dagger}$ is a 12-component spinor containing operators of the form 

\begin{align}
B_{\mathbf{k}}^{\dagger} &=\left(\begin{array}{cccc}
b_{\mathbf{k}A1}^{\dagger} & b_{\mathbf{k}B3}^{\dagger} & b_{-\mathbf{k},A1} & b_{-\mathbf{k},B3}\end{array}\right. \nonumber\\
 &\quad\quad\begin{array}{cccc}
b_{\mathbf{k}A2}^{\dagger} & b_{\mathbf{k}A4}^{\dagger} & b_{\mathbf{k}B2}^{\dagger} & b_{\mathbf{k}B4}^{\dagger}\end{array}\nonumber\\ 
 &\,\,\,\,\quad\left.\begin{array}{cccc}
b_{-\mathbf{k},2A} & b_{-\mathbf{k},A4} & b_{-\mathbf{k},B2} & b_{-\mathbf{k},B4}\end{array}\right),\label{eq:Bk}
\end{align}
and $\mathcal{H}_{\mathbf{k}}$ is a 12$\times$12 Hermitian matrix. The ordering of the spinor $B_{\mathbf{k}}^{\dagger}$ is motivated by the fact that bosons of colors $m=1$ and $m=3$ are decoupled from bosons with $m=2$ and $m=4$. This implies that $\mathcal{H}_{\mathbf{k}}$ can be written in a block diagonal form as
\begin{align}
\mathcal{H}_{\mathbf{k}}=\left(\begin{array}{ccc}
\mathcal{H}_{\mathbf{k}}^{\left(1,3\right)} & 0 & 0\\
0 & \mathcal{H}_{\mathbf{k}}^{\left(2,4\right)} & 0\\
0 & 0 & \left[\mathcal{H}_{-\mathbf{k}}^{\left(2,4\right)}\right]^{\ast}
\end{array}\right),
\end{align}
in which all the block matrices are 4$\times$4. Diagonalization of the LFWT Hamiltonian gives rise to six flavor 
dispersions $\omega_{\lambda}(\mathbf{k})$ that will be discussed below. We verify that the flavor waves originated from $\mathcal{H}_{\mathbf{k}}^{\left(2,4\right)}$ conserve the total number of bosons, 
in contrast to what happens for $\mathcal{H}_{\mathbf{k}}^{\left(1,3\right)}$. The constant term in the second and third lines of 
Eq. (\ref{eq:HLFWT}) gets canceled in the diagonalization and do not contribute to the ground state energy.

\begin{figure}
\begin{centering}
\includegraphics[width=0.47\columnwidth]{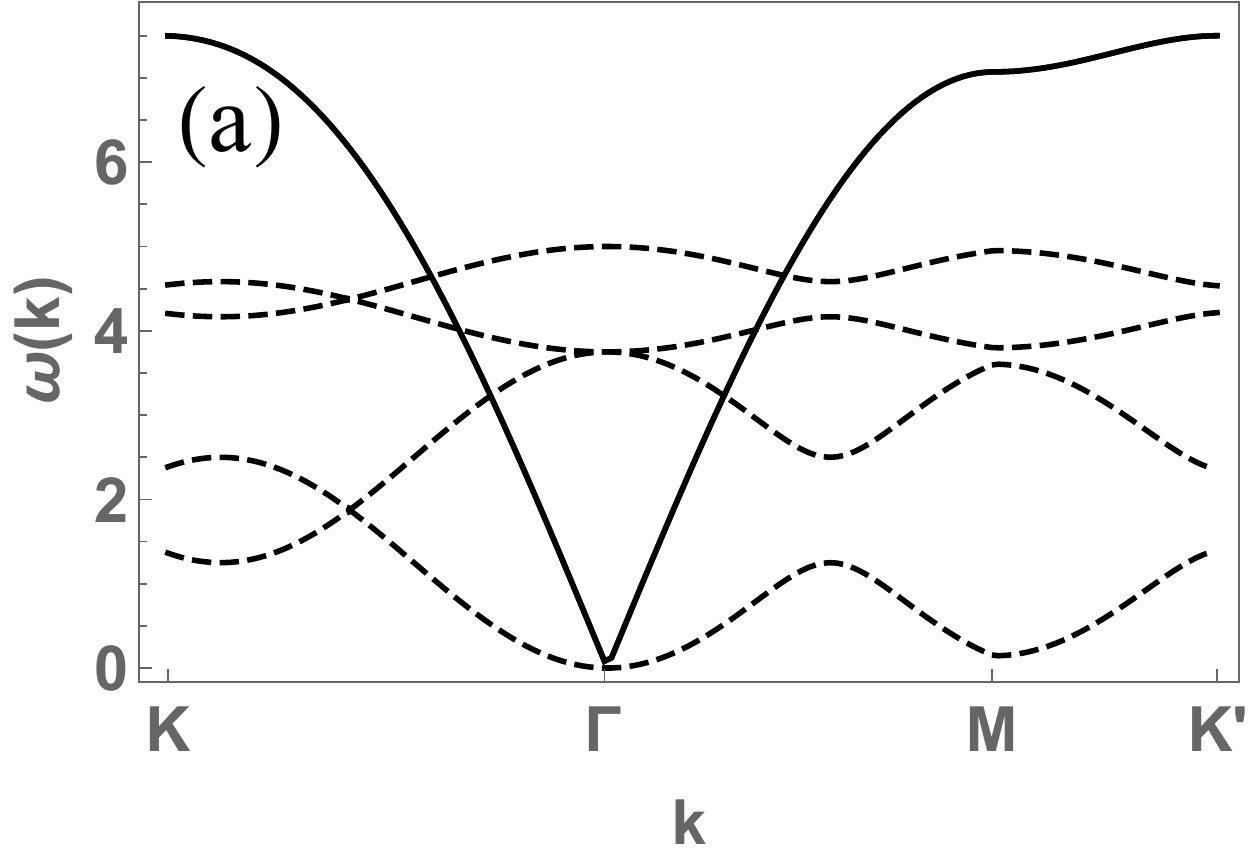}$\qquad$\includegraphics[width=0.47\columnwidth]{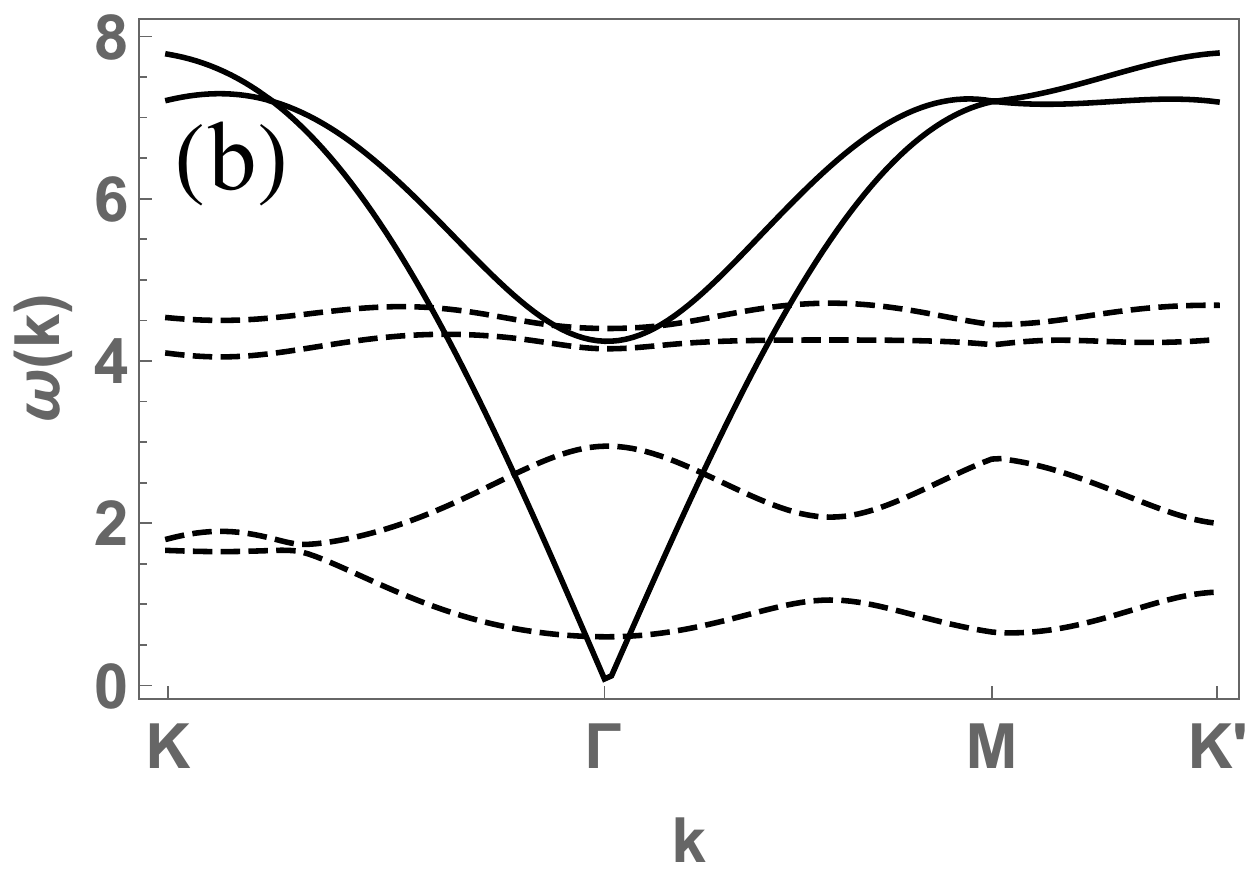}
\par\end{centering}
\begin{centering}
\includegraphics[width=0.47\columnwidth]{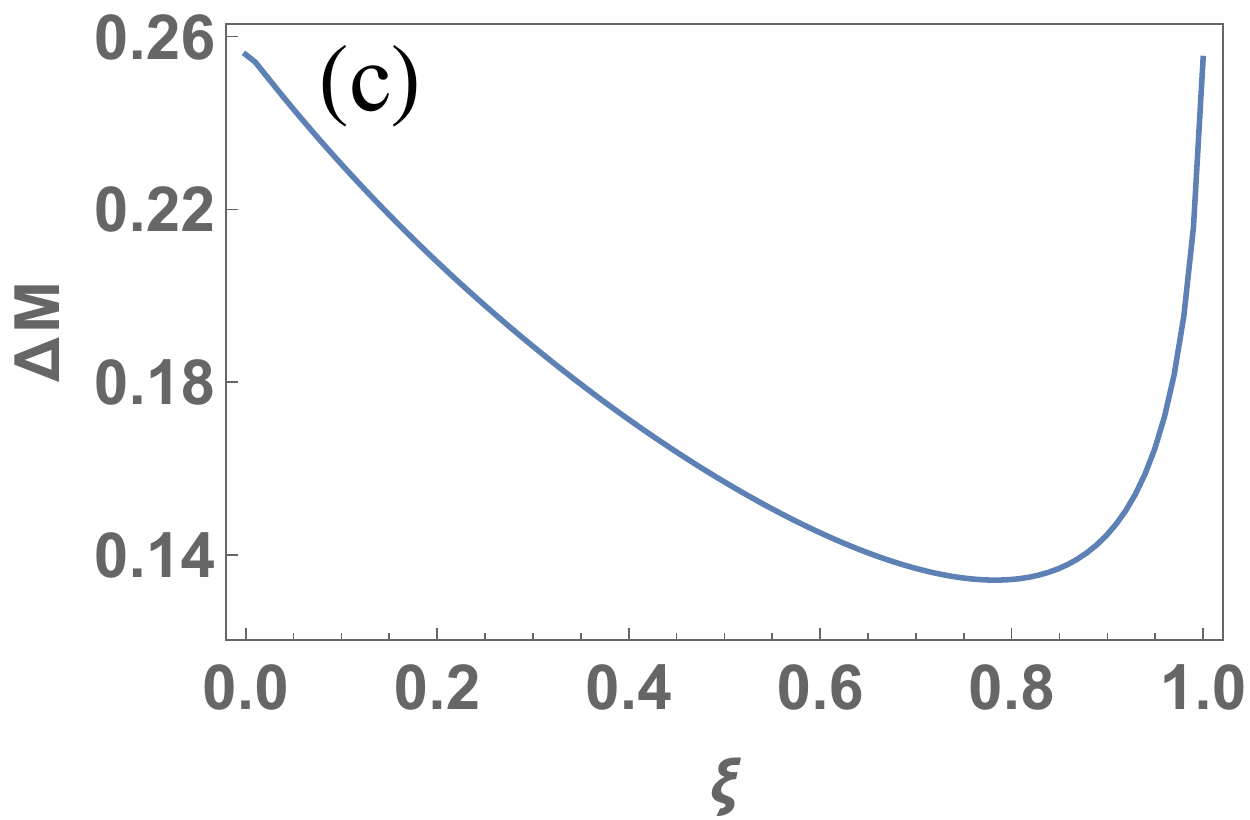}$\qquad$\includegraphics[width=0.47\columnwidth]{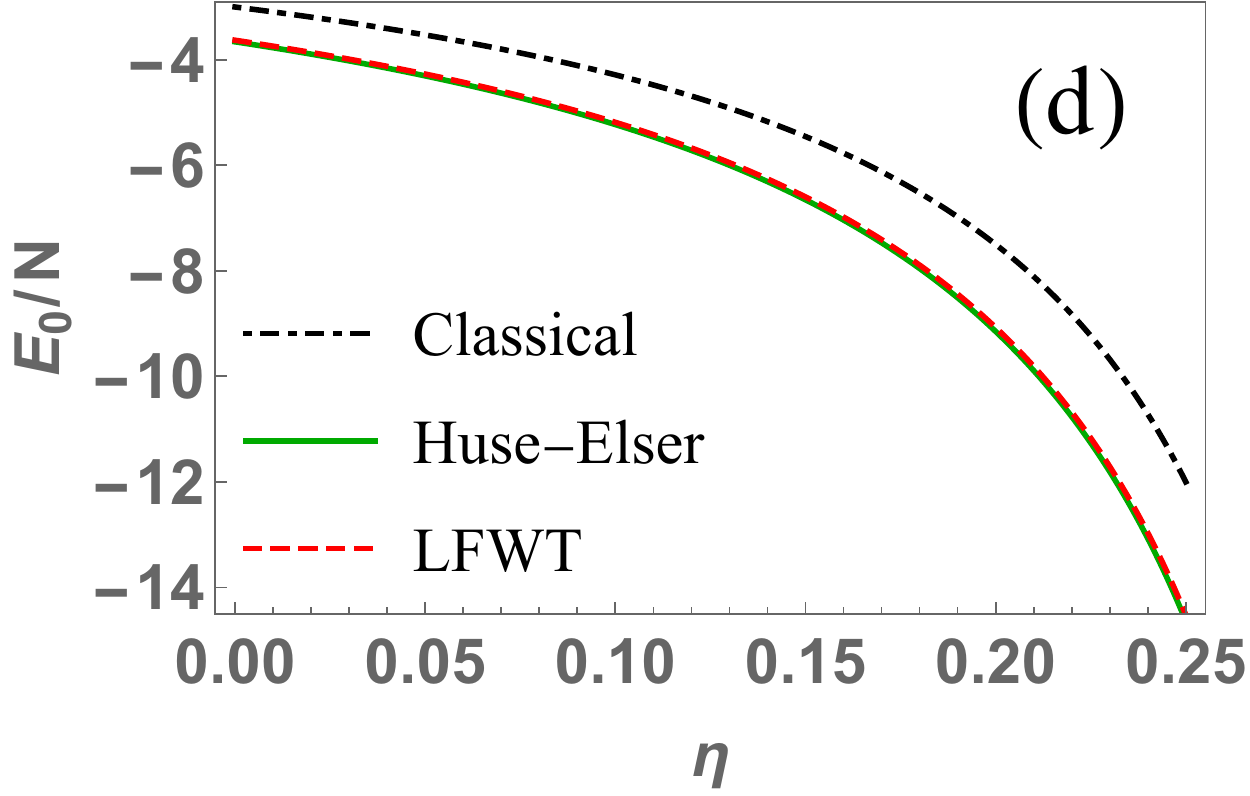}
\par\end{centering}
\caption{\label{fig:LFWTdispDS} (a) Dispersions $\omega_{\lambda}(\mathbf{k})$ for $\lambda = 1,2$ (continuous line) 
and $\lambda=3,...,6$ (dashed lines) of the ordered state FM AFOxz for $\eta=0.2$ and $\xi=0$. 
(b) Dispersions $\omega_{\lambda}(\mathbf{k})$ of the ordered state FM AFOxz 
for $\eta=0.2$ and $\xi=0.4$ with the same convention set in (a). 
(c) Correction to the order parameter $\Delta M$ as a function of $\xi$ for $\eta=0.2$
(d) Ground state energy as a function of $\eta$ at $\xi=0$ for the FM AFOxz
state comparing the classical energy, the LFWT energy, and the Huse-Elser
energy.} 
\end{figure}

Let us now turn to the flavor-wave dispersions $\omega_{\lambda}(\mathbf{k})$ with $\lambda=1,2$, which are related to $\mathcal{H}_{\mathbf{k}}^{\left(1,3\right)}$. Only the exchange constant $J_{1}$ appears in this sector of the LFWT Hamiltonian as a global multiplicative factor. Therefore, the shape of the dispersions $\omega_{\lambda}(\mathbf{k})$ does not vary with $\eta$ and the bandwidth is directly proportional to $J_{1}$. Figure \ref{fig:LFWTdispDS}(b) shows $\omega_{\lambda}(\mathbf{k})$ for $\xi=0$ in solid lines. In this case, we observe two degenerate bands with linear dispersion at the $\Gamma$ point. This degeneracy is lifted by bond-dependent interactions as shown in Fig. \ref{fig:LFWTdispDS}(b). The resulting band retains a Goldstone mode and another gapped mode. 

We now turn to stability criteria given by $\omega_{\lambda}(\mathbf{k})$ with $\lambda = 3,...,6$ 
obtained from $\mathcal{H}_{\mathbf{k}}^{\left(2,4\right)}$ (see dashed lines in Figs. \ref{fig:LFWTdispDS}(a) and (b)). First, these bands 
become zero-energy flat bands in the limit $(\xi,\eta)\rightarrow (0,0)$. This provides another indication of the instability 
of the ordered state at the SU(4)-symmetric point. Second, $\omega_{\lambda}(\mathbf{k})$ become negative
depending on the values of $\eta$ and $t^{\prime}$, which provides yet another instability flag. The region in which 
this form of instability disrupts an otherwise favored FM AFOxz phase was found numerically and is indicated in gray in  
Fig. \ref{fig:phase_diagram}. Reference \citep{Smerald2014} also encountered an unidentified phase with ED for 
small clusters in a close region of the parameter space. LFWT suggests that such phase still exists 
in the thermodynamic limit, but it is incapable of diagnosing its characteristics. 
When these bosonic modes display strictly positive frequencies, they do not alter
the LFWT ground state. Thus, they do not affect the reduction of the order parameter
nor the zero-point energy. The energy of the ordered state calculated with LFWT, $E_{\text{LFWT}}$, 
considers then only the integration of the modes $\lambda=1,2$ and is given by:

\begin{align}
\frac{E_{\text{LFWT}}}{N}=-2J_{1}\left(3+\xi^{2}\right)+\frac{1}{N}\sum_{\mathbf{k}}\sum_{\lambda=1}^{2}\omega_{\lambda}\left(\mathbf{k}\right).\label{eq:ELFWT}
\end{align}

The correction to the order parameter is provided by Fig. \ref{fig:LFWTdispDS}(c),
which shows that $\Delta M\sim0.14-0.25$ when $\xi<0.99$.
We observe a divergence of $\Delta M$ as $\xi\rightarrow 1$. This is consistent
with the mean-field phase transition occurring at this point due to
the six-sublattice mapping discussed above, see Fig. \ref{fig:Classical_diagram}.
Away from this line, the ordered phase FM AFOxz acquires only 
mild corrections to the order parameter. 

To further check the feasibility of the LFWT energies, we also construct
a variational wave function for the ordered FM AFOxz phase, along
the line $\xi=0$, following the proposal by Huse and
Elser \citep{Huse1988}. Using standard VMC techniques \citep{Huse1988,Ferrari2017,Gros1989},
we then find the ground state energy as a function of $\eta$, shown
in Fig. \ref{fig:LFWTdispDS}(d). The Huse-Elser energies display
a remarkable agreement with the LFWT theory, with only a slightly lower energy. 
We can then argue that LFWT and variational methods provide consistent results
for the FM AFOxz energy, which allows the use of $E_{\text{LFWT}}$ as the estimator for the ordered state energy.

\section{Quantum Spin-Orbital Liquid and valence bond crystals\label{sec:QSOL_VBC}}

In Ref. \cite{Corboz2012}, a QSOL was  identified as the best  candidate for the ground state of the SU(4) Heisenberg model. Within parton mean-field theory \cite{Savary2016}, the  state can be pictured as four flavors of free fermions hopping in a background with $\pi$ flux of the emergent gauge field through every hexagon of the lattice. With the constraint of one fermion per site,   quarter filling of the bands gives rise to a gapless spectrum with a Dirac dispersion at low energies. Being gapless, such two-dimensional QSOL is in principle stable beyond the mean-field level, when gauge fluctuations are included \cite{Hermele2004}.

An important question is whether such QSOL survives in the presence of SU(4)-symmetry-breaking perturbations like the ones considered in model \eqref{eq:complete_Hamiltonian}. Based on exact diagonalization on small clusters, Ref. \cite{Smerald2014} argued for a QSOL phase over an extended region in the parameter space around the SU(4)-symmetric point. In the following, we use VMC methods to investigate the stability of the QSOL in our model. In contrast to ED, the computational time to obtain an observable mean-value
and variance with VMC increases polynomially instead of exponentially. Hence,
VMC calculations can then be performed in larger samples and allows a more reliable extrapolation to the
thermodynamic limit. Moreover, VMC algorithms can be used to
study VBC states parting from small modifications of QSOL wave functions,
making it an adequate technique to evaluate the energetics of these two
classes of states.

\subsection{Quantum Spin-Orbital Liquid\label{subsec:Parton}}

First, we introduce the fermionic parton representation of the SU(4) generators
\begin{equation}
S_{a}^{b}(i)=f_{i,a}^{\dagger}f^{\phantom\dagger}_{i,b},\label{eq:partonSab}
\end{equation}
 in which $a=1,...,4$
labels the color states and $i$ labels the lattice site. The fermionic operators
satisfy the canonical relation $\{ f_{i,a},f_{j,b}^{\dagger}\} =\delta_{ij}\delta_{ab}$ and define
a Fock space. Equation (\ref{eq:partonSab}) is not an exact rewriting of
$S_{a}^{b}(i)$ since the physical Hilbert space is isomorphic only
to the Fock subspace that satisfies the local single-occupancy constraint $ \sum_af^\dagger_{i,a}f^{\phantom\dagger}_{i,a}=1$.
VMC allows an evaluation of averages for observables after implementing
a numerical projection to the physical space.

\begin{figure}
\begin{centering}
\includegraphics[width=0.54\columnwidth]{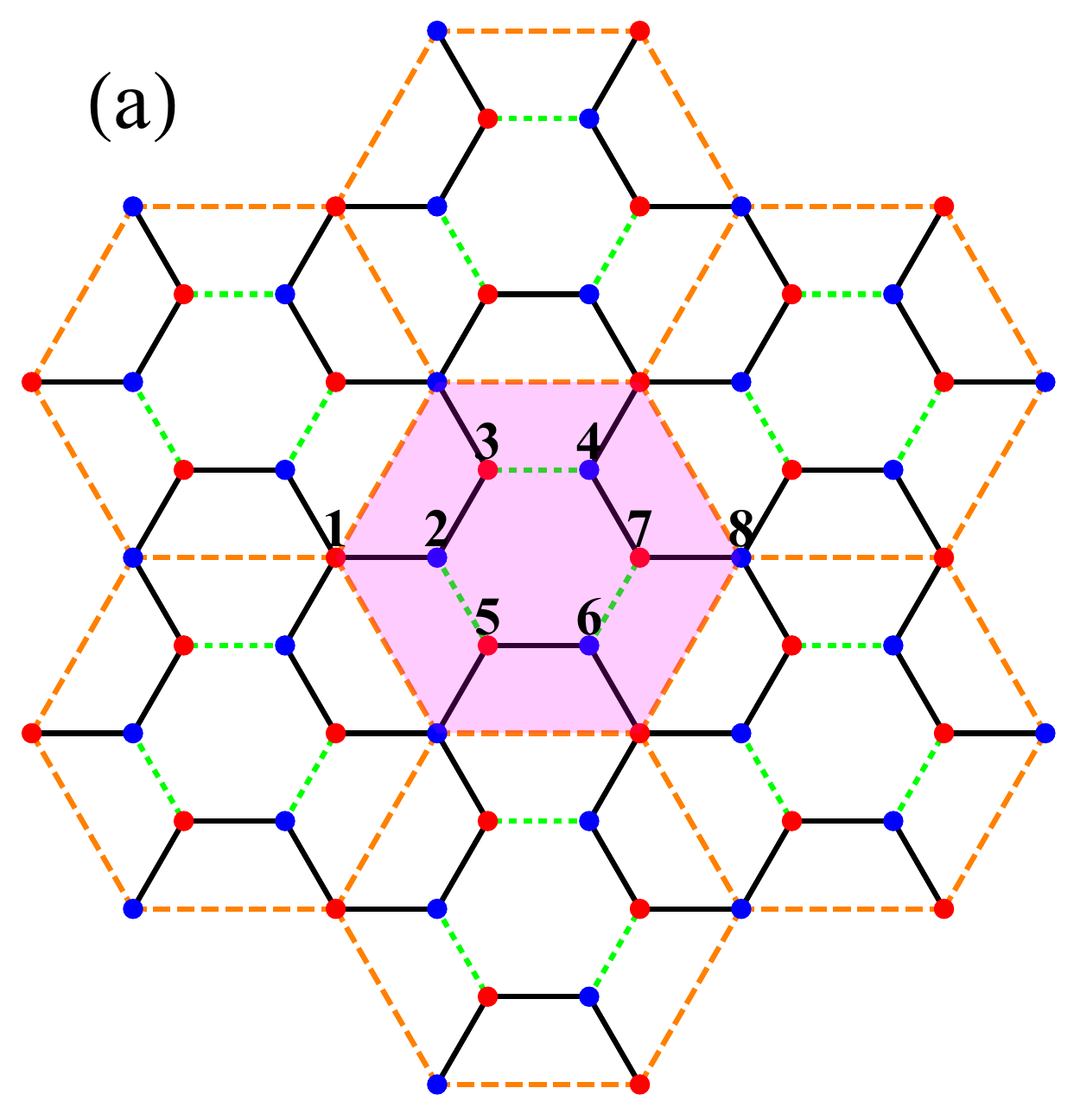}$\,$\includegraphics[width=0.38\columnwidth]{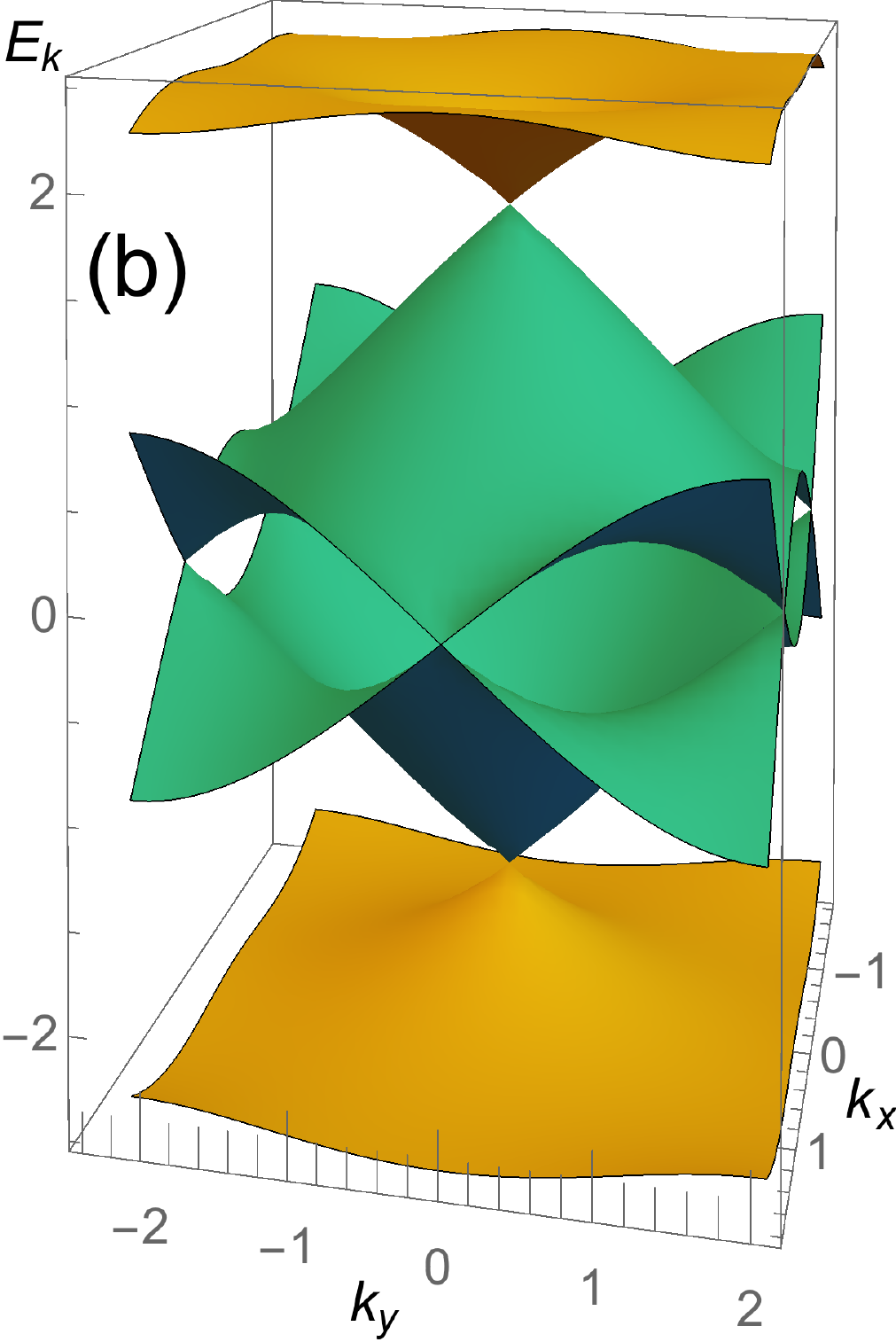} 
\par\end{centering}
\caption{\label{fig:MFTdispersion} (a) Mean-field ansatz for the $\pi$-flux state. The links in black have $\chi_{ij}=\chi$  ($\phi_{ij}=0$) while the links in green have $\chi_{ij}=-\chi$ ($\phi_{ij}=\pi$). (b) Dispersion relation of the $\pi$-flux state. The parton mean-field ground state is obtained through the occupation of the states in the lowest energy band displayed in yellow. }
\end{figure}

To generate a trial wave function for the QSOL, we first determine the ground state of the mean-field Hamiltonian
\begin{equation}
\mathcal{H}_{\text{mf}}=-\sum_{a}\sum_{\langle ij\rangle}\left[\chi_{ij}f_{i,a}^{\dagger}f^{\phantom\dagger}_{j,a}+\text{h.c.}\right],\label{eq:Hmf0}
\end{equation}
where the  choice of parameters $\chi_{ij}\in \mathbb C$ specify the mean-field ansatz. This ansatz is invariant under SU(4) transformations, thus enforcing a higher symmetry on the state than the SU(2)$\times\mathbb Z_3$ symmetry of model \eqref{eq:complete_Hamiltonian}. Translational invariance requires the absolute value of $\chi_{ij}$ to be uniform: $\chi_{ij}=\chi e^{i\phi_{ij}}$, where $\chi>0$ and $\phi_{ij}$ is the phase associated with the link $\langle ij \rangle$.  The gauge flux  $\Phi$ on each elementary hexagonal plaquette is defined by $e^{i\Phi}\equiv\prod_{\langle ij\rangle\in \hexagon}e^{i\phi_{ij}}$. Here we focus on the $\pi$-flux state  with $\Phi=\pi$ through every hexagon \cite{Corboz2012} (see Fig. \ref{fig:MFTdispersion}(a)). The corresponding dispersion relation showing a single Dirac cone at the
$\Gamma$ point is illustrated in Fig. \ref{fig:MFTdispersion}(b). 

\begin{figure}
\begin{centering}
\includegraphics[width=0.8\columnwidth]{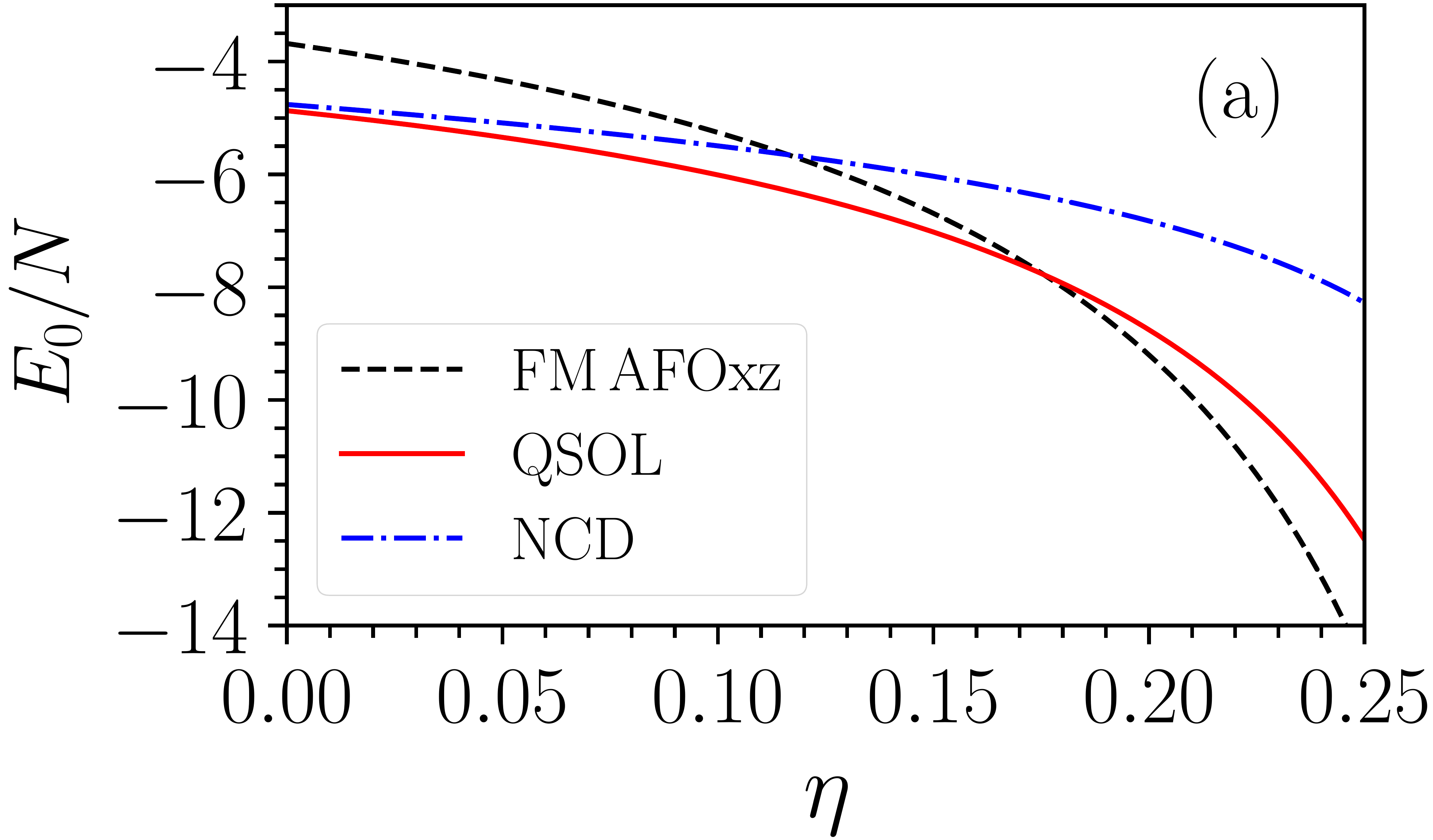}
\par\end{centering}
\begin{centering}
\includegraphics[width=0.8\columnwidth]{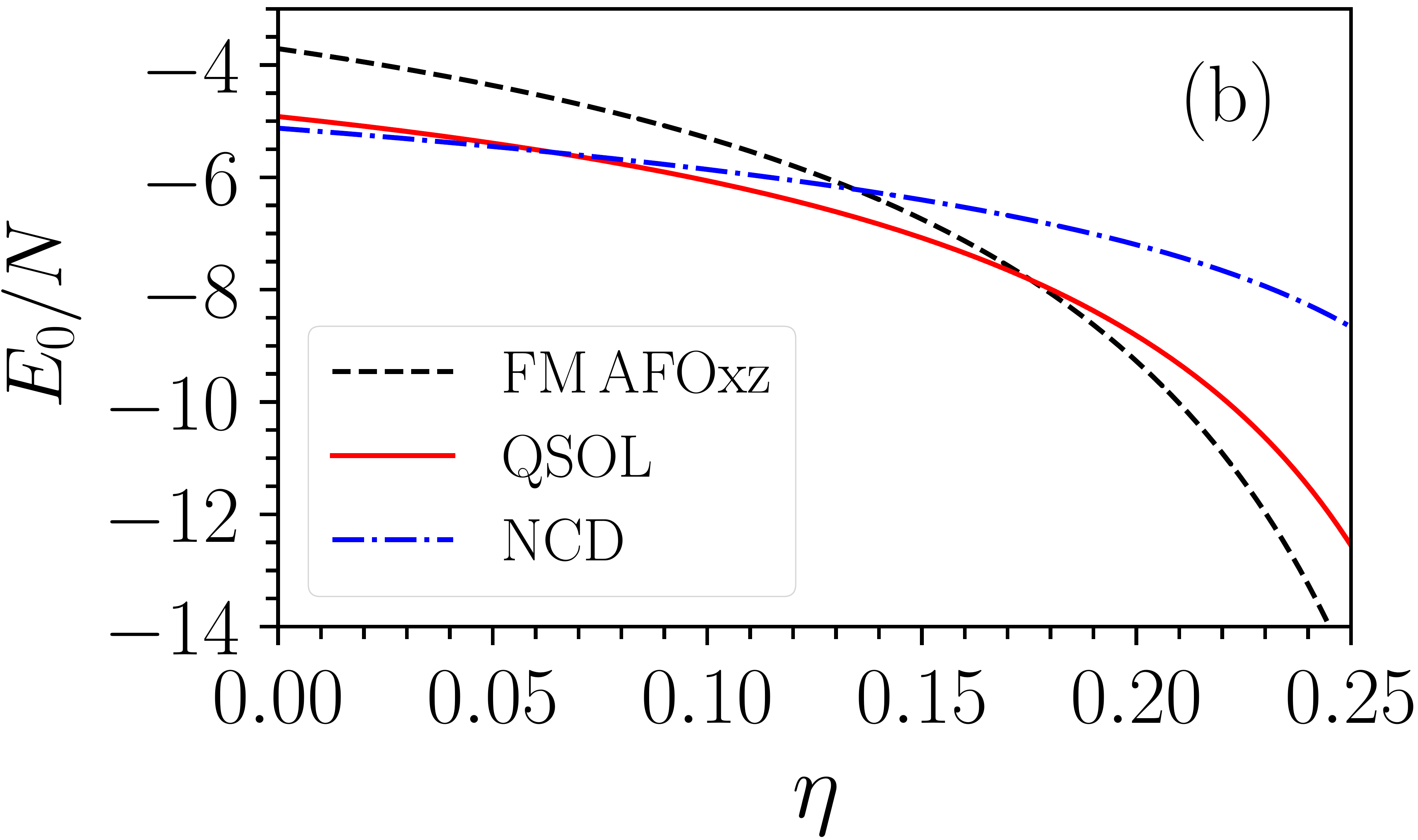}
\par\end{centering}
\caption{\label{fig:detail_diagram} Comparison of the energy, per site, of
three different states: FM AFOxz, QSOL, and NCD (see text) as a function
of the Hund's coupling $\eta$ for (a) $\xi=0.2$ and (b) $\xi=0.25$.
We identify the phase transitions as the crossing points between the
different curves.}
\end{figure}

The ground state at the mean-field level $|\psi_0\rangle$ is obtained by 
filling the lower band shown in Fig. \ref{fig:MFTdispersion}(b). 
We just outline the VMC procedure, since technical details on how to perform
the Gutzwiller projection of the mean-field wave functions
can be found in the specialized literature \citep{Gros1989, Corboz2012,Natori2016,Natori2018}. 
The energy of the Gutzwiller-projected $|\psi_0\rangle$ is calculated for the
KK model in Eq. (\ref{eq:complete_Hamiltonian-1}) for different values of 
$\xi$ and $\eta$ (see Fig. \ref{fig:detail_diagram}). We consider honeycomb lattices of
linear length $L$ and $N=2L^{2}$ sites with $L=6$, $12$, and $18$.
An initial state for the Monte Carlo evaluation is chosen by randomly 
placing each color at $N/4$ sites of our lattice. Our Monte Carlo move consists in exchanging 
a random pair of sites containing distinct colors, which is accepted or rejected according to the
general Metropolis algorithm. A Monte Carlo sweep consists of $\sim10^{3}$ exchanges
attempts. After every sweep, we compute the ground state energy $E_{0}$. 
We typically perform $\sim10^{5}$ sweeps, with half of the steps discarded for equilibration.

We compared the energy of this particular QSOL to that of the ordered
state FM AFOxz. Recall that, as discussed in Sec. \ref{sec:Ordered-States},  
the AFM AFOxz phase is unstable against quantum fluctuations and disappears completely.
We find that the QSOL extends itself away from the SU(4) point, and
covers an appreciable portion of the phase diagram before giving room
for the FM AFOxz at $\eta\approx0.175$, a value which is essentially
independent of $\xi$. The presence of the FM AFOxz phase at
large $\eta$ is expected: Hund's coupling favors a ferromagnetic
spin alignment, while the local Hubbard repulsion favors a staggered
orbital occupation \citep{khomskii2014transition}. Nevertheless,
the QSOL originally identified in Ref. \citep{Corboz2012} survives
the introduction of a finite Hund's coupling and orbital anisotropy,
and it is a competitive ground state for \textcolor{black}{KK models
in the honeycomb lattice.}

\subsection{Valence bond crystals \label{subsec:VMC-VBCs}}

\begin{figure}
\begin{centering}
\subfloat{\includegraphics[width=0.3\columnwidth]{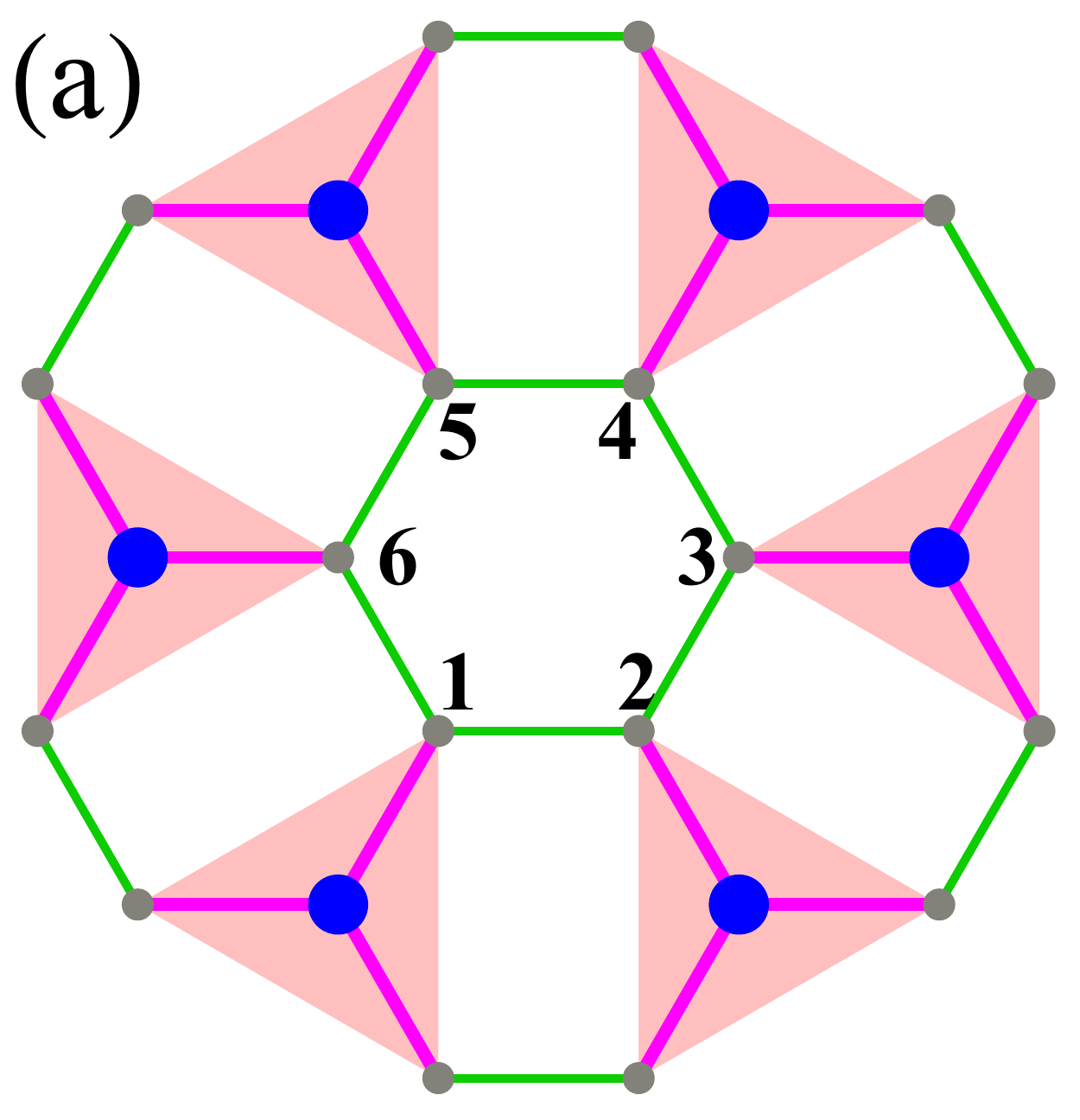}}\quad\subfloat{\includegraphics[width=0.3\columnwidth]{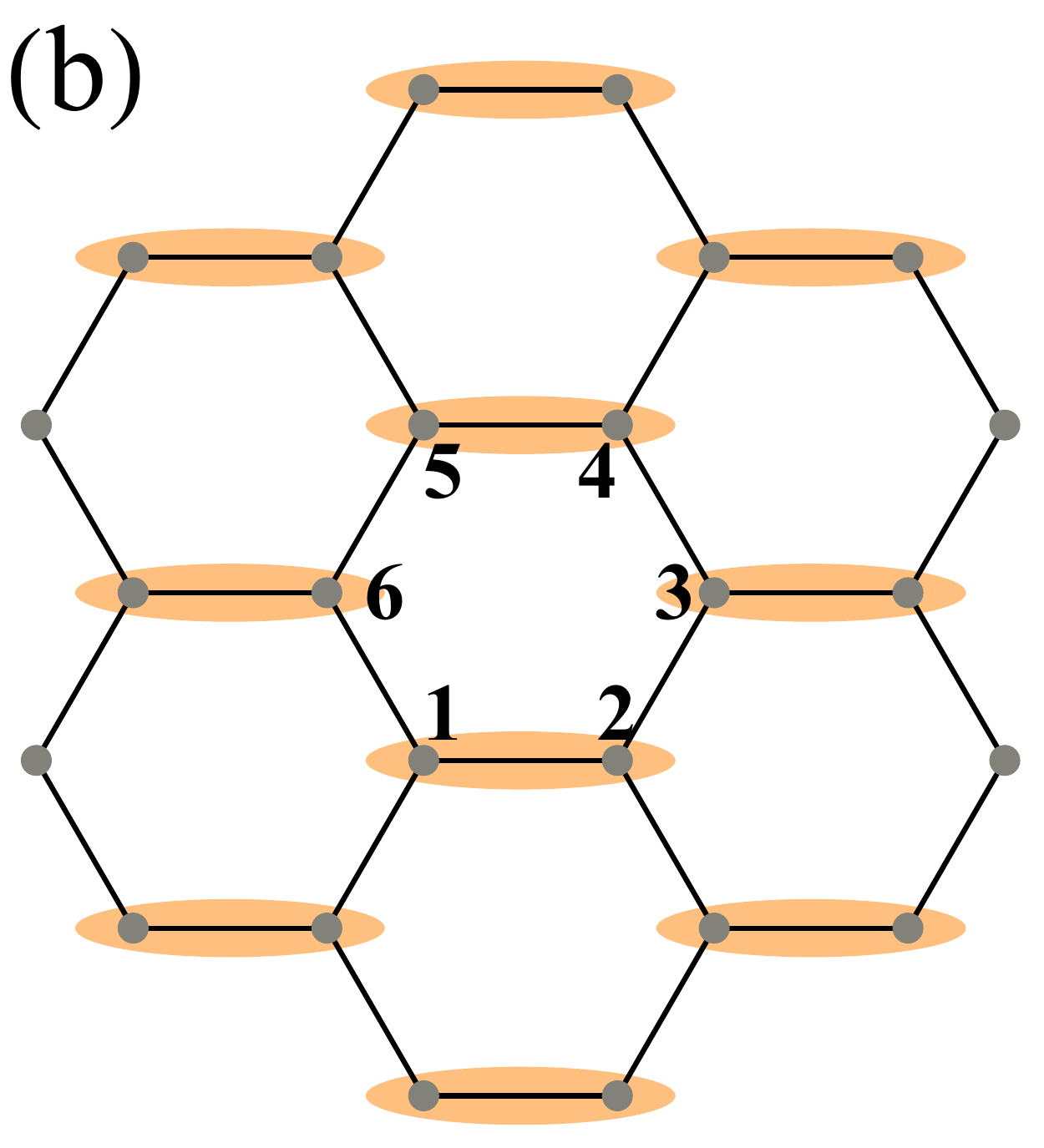}}\quad\subfloat{\includegraphics[width=0.3\columnwidth]{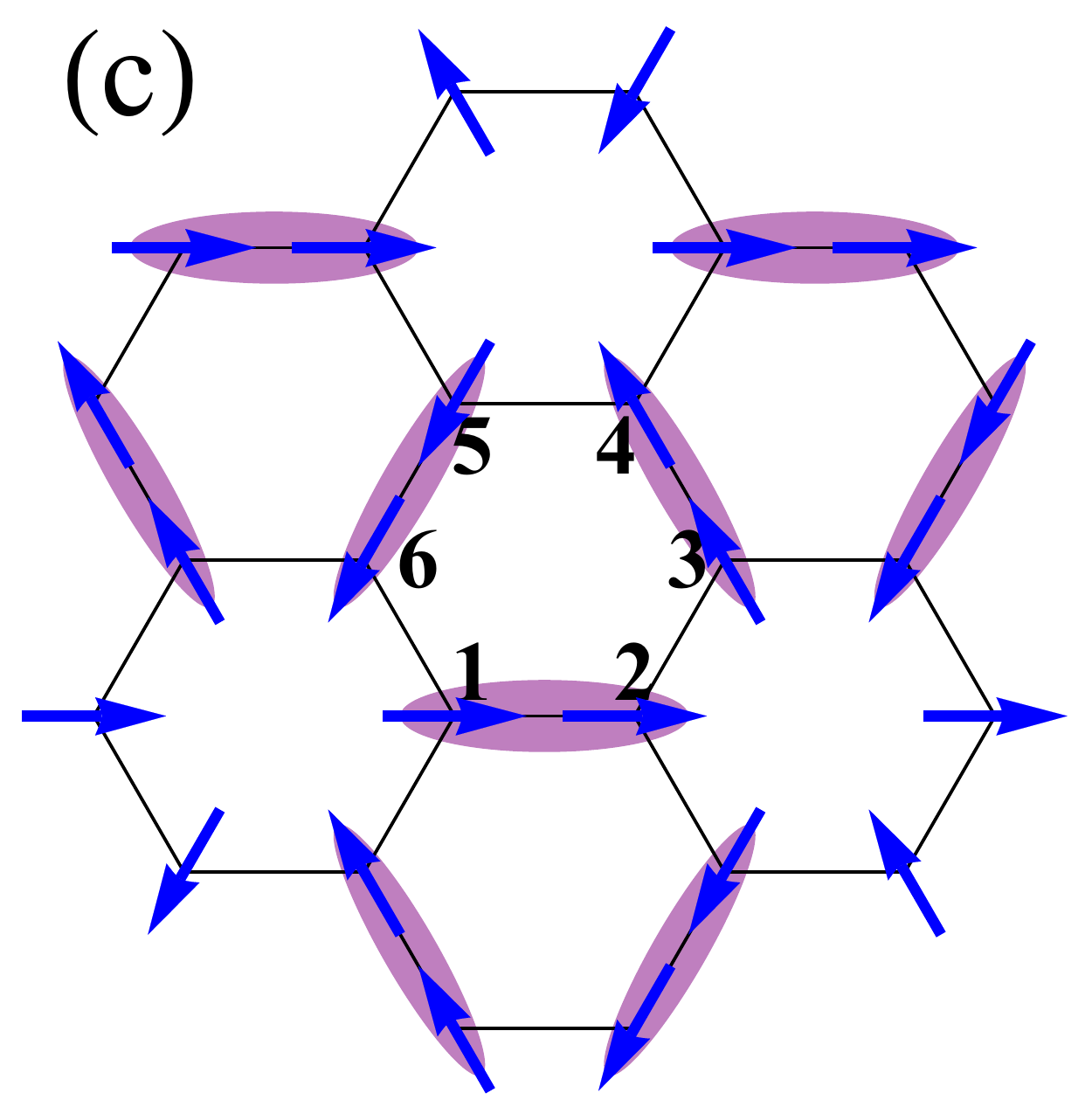}}
\par\end{centering}
\caption{\label{fig:trivial_paramagnets} Valence bond crystal states considered in this work:   (a)
 tetramerized state, in which the sites inside magenta triangles form an SU(4) singlet; (b)  dimerized state
given by the product of a ferromagnetic order and collinear orbital dimers; (c)  product state of  non-collinear
spin dimers and the  ferro-orbital vortex state of Fig. \ref{fig:Classical_diagram}(d).}
\end{figure}

\begin{figure}
\begin{centering}
\includegraphics[width=0.48\columnwidth]{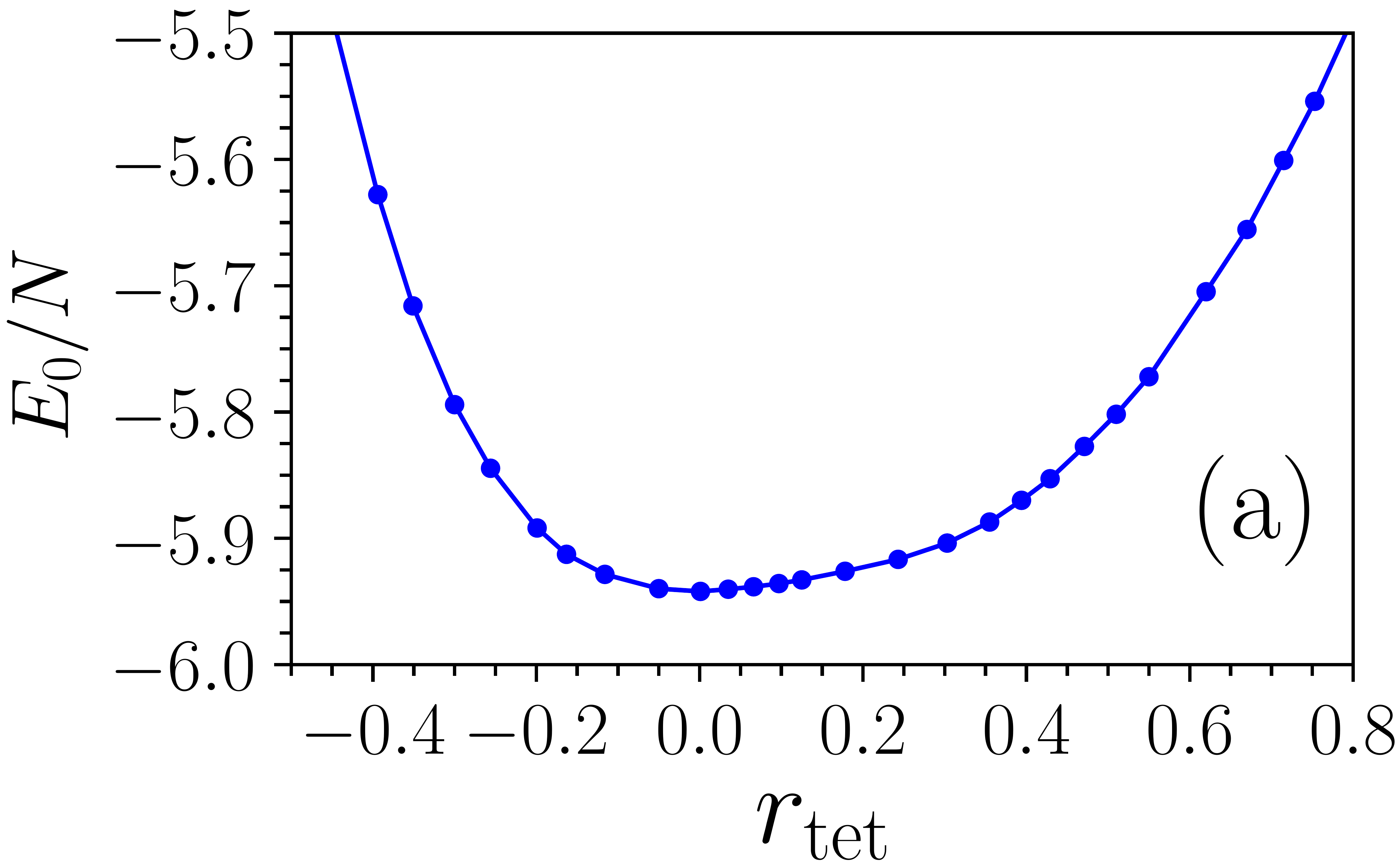}\includegraphics[width=0.48\columnwidth]{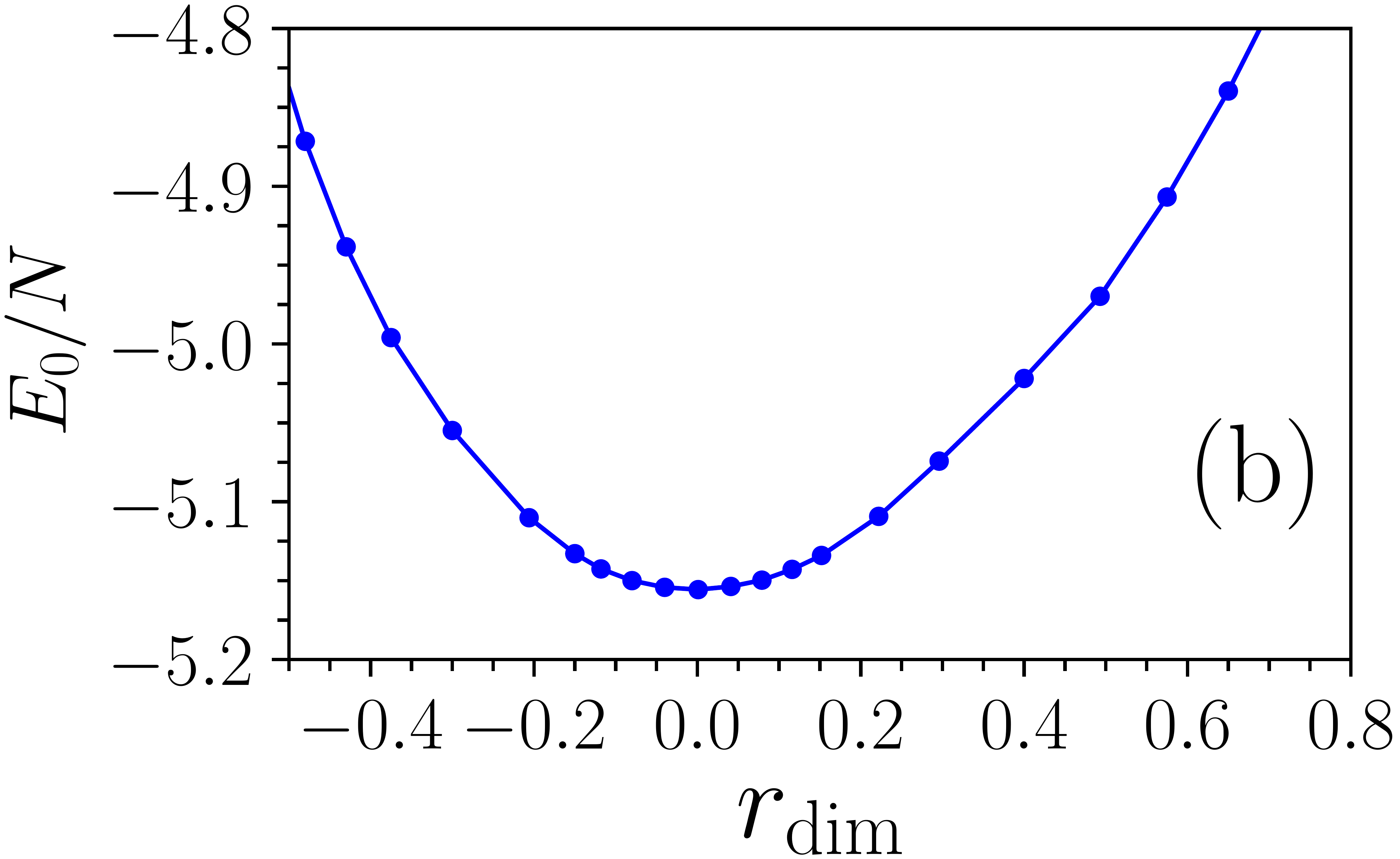}
\par\end{centering}
\caption{\label{fig:tetra}(a) Ground state energy of the tetramerized
state, per site, as a function of the tetramerization parameter $r_{{\rm tet}}$.
(b) Ground state energy of the dimerized state, per site, as a
function of the dimerization parameter $r_{{\rm dim}}$. We considered
$\xi=\eta=0.1$ for both curves.}
\end{figure}

In  Section \ref{subsec:Parton}, we found that the QSOL is eventually replaced
by an ordered state for large enough  Hund's coupling $\eta$.
Now we want to investigate different instabilities of the QSOL, specially
as a function of $\xi$ and focused on the formation of VBC states. We start this investigation
with the tetramerized state. Here the spins form four-site singlet plaquettes breaking translational
symmetry but preserving the SU(4) symmetry \citep{Lajko2013}. A possible
tetramer covering of the honeycomb lattice is illustrated in Fig.
\ref{fig:trivial_paramagnets}(a). We tested the stability of
the $\pi$-flux state against this tetramerization pattern by considering
variational wave functions generated by the mean-field Hamiltonian 

\begin{equation}
\mathcal{H}_{{\rm mf,t}}=\sum_{m=1}^{4}\left[\sum_{i}\tilde{\varepsilon}_{i}f_{im}^{\dagger}f_{im}-\sum_{\left\langle ij\right\rangle }\left(\tilde{\chi}_{ij}f_{im}^{\dagger}f_{jm}+{\rm h.c.}\right)\right].\label{eq:h_tetra}
\end{equation}
Here we keep the $\pi$-flux ansatz, so we
modulate the sign of $\tilde{\chi}_{ij}$ as in Fig. \ref{fig:MFTdispersion}(d), but we also allow for non-uniform magnitude of the mean-field parameters: $|\tilde{\chi}_{ij}|=\chi^{{\rm tet}}$ if sites $i$ and
$j$ belong to the same tetramer and $|\tilde{\chi}_{ij}|=\chi$
otherwise. Furthermore, we define a negative on-site energy $\tilde{\varepsilon}_{i}$ for sites at the center of the tetramers 
(see sites highlighted in Fig. \ref{fig:trivial_paramagnets}(a)). For $\tilde{\varepsilon}_{i}=0$ and $\chi_{ij}^{{\rm tet}}=\chi$,
we recover the uniform $\pi$-flux state. The fully tetramerized state
is the product of independent four-site SU(4) singlets throughout
the lattice \citep{Li1998}. To quantify the degree of tetramerization
of the projected wave functions, we consider the permutation operator
between nearest neighbors: $P_{ij}=\sum_{a,b}S_{a}^{b}(i)S_{b}^{a}(j)$,
with the color exchange operators $S_{a}^{b}(i)$ defined in Eq. (\ref{eq:Sab}).
The tetramerization order parameter is defined by \citep{Lajko2013}
\begin{equation}
r_{\text{tet}}=\frac{4}{5}\left(P_{1}-P_{2}\right),\label{eq:rtet}
\end{equation}
where $P_{1}$ is the expectation value of $P_{ij}$ for bonds connecting
sites inside a tetramer, while $P_{2}$ is the average of $P_{ij}$
for any other bond (Fig. \ref{fig:trivial_paramagnets}(a)). The parameter
$r_{{\rm tet}}\left(\tilde{\varepsilon},\,\tilde{\chi}\right)$ is
normalized such that $r_{{\rm tet}}=1$ in the four-site plaquette
product state. For each value of $\tilde{\varepsilon}$, we select
the value of $\tilde{\chi}=\tilde{\chi}_{{\rm min}}\left(\tilde{\varepsilon}\right)$
that gives the lowest energy within VMC and compute the corresponding
tetramerization order parameter $r_{{\rm tet}}\left(\tilde{\varepsilon},\,\tilde{\chi}_{{\rm min}}\left(\tilde{\varepsilon}\right)\right)$.
Our VMC results in Fig. \ref{fig:tetra}(a) illustrate that the lowest
energy is obtained for $r_{{\rm tet}}=0$ and thus the uniform state
is always selected in the region where the QSOL is stable. This implies
that the QSOL is stable against tetramerization, in accordance with
the results of Ref. \citep{Lajko2013} at the SU(4) point.

Next, we consider valence bond crystals constructed after a mean-field decoupling of   spin and orbital degrees of freedom 
\citep{Smerald2014}. Since it neglects spin-orbital entanglement, this approximation should break down close to the SU(4) point. Nevertheless,
it allows us to search for other trial states which may
be stable, for instance, in the region $\xi\sim 1$.  

As a first example, we consider that the spins are fully polarized while the orbitals form  the  collinear dimer pattern  shown in Fig. \ref{fig:trivial_paramagnets}(b).
A simple product state of orbital dimers produces the following ground
state energy: $E_{0}/N=-3(1+\xi^2)J_{1}$. It is interesting
that this energy improves as one moves away from the SU(4)   point, 
in accordance with our general discussion. We also performed
a full VMC study with a mean-field Hamiltonian similar to the one
in Eq. (\ref{eq:h_tetra}). The dimerization order parameter is $r_{{\rm dim}}=\tilde{P}_{1}-\tilde{P}_{2}$,
with $\tilde{P}_{1}$ the average value of the \emph{two-color} permutation
operator on bonds forming a dimer, and $\tilde{P}_{2}$ is the average
value for any other bond. By definition, $r_{{\rm dim}}=1$ in the
product state discussed previously. Our VMC results in Fig. \ref{fig:tetra}(b)
highlight that the lowest energy  occurs for $r_{{\rm dim}}=0$
and thus this orbital dimerization is never favored. Moreover, the overall energy
is not competitive and this state does not appear in the phase
diagram.

Now we assume that the spins dimerize, while the orbitals develop some type of classical order. A spin dimer is clearly
 not favored by Hund's coupling and is most likely to be present
at small $\eta$. In this limit, a reasonable guess for the orbital
dependence would be that orbitals belonging to a spin dimer are parallel.
However, ferro-orbital order is too high in energy in the region $t^{\prime}\lesssim t$.
 Another more promising choice is the ferro-orbital vortex state  
in Fig. \ref{fig:Classical_diagram}(d), coupled to a ``Kekule'' arrangement of nearest-neighbor spin dimers \citep{Xu2018b}. The resulting spin-orbital 
state is depicted  in Fig. \ref{fig:trivial_paramagnets}(c). We follow the nomenclature of Ref. \citep{Smerald2014} and refer
to it as a noncollinear spin dimers (NCD) phase. To test this state
within VMC, we assume ferro-orbital vortex order and perform a VMC calculation
in the resulting spin Hamiltonian allowing for     spin dimerization,
following the same approach as described above. We now find $r_{{\rm dim}}\neq0$
in the NCD state. Importantly, this is the best variational state
in the region $\xi\gtrsim0.2$ and $\eta\lesssim0.2$, as indicated by Fig.
\ref{fig:detail_diagram}. The detailed comparison of the energies of the variational states FM AFOxz, QSOL, and NCD leads to the phase diagram in Fig. \ref{fig:phase_diagram}. However, we find no competitive candidate, within VMC, for the LFWT unstable region in Fig. \ref{fig:phase_diagram}. In particular, we investigated states with partially polarized spins, following the suggestion of Ref. \citep{Smerald2014}, but their energy is never competitive (see, for instance, Fig. \ref{fig:tetra} (b)).

\section{Discussion\label{sec:Discussion}}

We revisited a Kugel-Khomskii  model in the honeycomb lattice previously studied in the context of 
spin-orbital physics of Ba$_3$CuSb$_{2}$O$_{9}$ and quarter-filled  twisted bilayer graphene \citep{Smerald2014,Venderbos2018}. 
This model contains an SU(4)-symmetric point at which a QSOL phase may be realized   \citep{Corboz2012}. Using a combination of 
analytical and numerical  techniques, we found that this QSOL covers an extended parameter regime in the phase diagram where we 
include the effects of Hund's coupling and bond-dependent frustrated exchange interactions. This result raises hopes 
that a QSOL state may be observed in honeycomb lattice materials with active spin and orbital degrees of freedom.

Ba$_3$CuSb$_{2}$O$_{9}$ contains Cu$^{2+}$ ions with a $3d^{9}$
configuration. In a first approximation, one may assume that this hole
has a fourfold degeneracy: a twofold spin degeneracy and a twofold
orbital degeneracy of the $e_{g}$ orbitals. Normally, one would expect
this degeneracy to be lifted and long-range order to develop for both spin and
orbital degrees of freedom at low temperatures. However, no spin freezing
is detected down to $20$ mK \citep{Quilliam2012}, considerably below
the Curie-Weiss temperature of $50$ K, and no evidence for a cooperative
Jahn-Teller effect is found down to $12$ K \citep{Nakatsuji2012,Katayama2015}.
These experimental observations motivated the proposal of this material
as a QSOL candidate \citep{Quilliam2012,Nakatsuji2012}. Nevertheless,
as stressed by Ref. \citep{Smerald2014}, the microscopic model in
Eq. (\ref{eq:complete_Hamiltonian}) is too simplistic to describe
Ba$_3$CuSb$_{2}$O$_{9}$ and a QSOL is likely
not its ground state. For this material, a more realistic Hamiltonian 
on a decorated honeycomb lattice should be taken into account.

In TBG, the orbital degrees of freedom originate from the two
Dirac points in the original Brillouin zone of each graphene sheet, 
which should be centered on a honeycomb superlattice due to symmetry
constraints \citep{Kang2018a,Koshino2018,Po2018}. The effective Hamiltonian 
obtained at quarter filling is the SU(4) Heisenberg model  \citep{Venderbos2018}. 
As we increase the Hund's coupling, we find long-range ferromagnetic
order in the spins and antiferromagnetic order in the orbitals, providing
a possible connection with a recently found spin-polarized state \citep{Seo2019,Sharpe2019,Serlin19}.
While longer-range exchange couplings are likely to be relevant in the Mott insulating phase of 
TBG \citep{Kang2019,Xu2018b}, an intriguing  possibility is that a spin-polarized phase exists in 
proximity to a QSOL \citep{Kiese2019} in this highly tunable system.

We close this paper with remarks about three solid-state platforms that would be 
described by similar KK models: the trilayer graphene/hexagonal 
boron nitride heterostructures (TLG/hBN) \cite{Classen2019,Schrade2019,Wu2019,ZhangB2019},
the tiny-angle TBG system \citep{Ramires2018} and the $j=3/2$ compound 
$\alpha$-ZrCl$_3$ \cite{Yamada2018}. KK models for TLG/hBN also display twofold orbital degeneracy,
but these Wannier orbitals are located on a triangular lattice, which implies that our results are not 
extendable to this system. However, our methodology is undoubtedly applicable to these models 
and can provide complementary results. The same comment applies to the tiny-angle
TBG under an electric field, which is possibly described by a KK model
on an emergent kagome lattice \citep{Ramires2018}. 
Concerning the layered honeycomb material $\alpha$-ZrCl$_3$, 
it is expected that extended versions of the minimal model derived in Refs. \citep{Yamada2018,Natori2018} 
would lead to exchange frustration similar to the ones discussed in this paper. 
The phase diagram of a realistic model for this compound would then present
extended regions of stability for the QSOL and the NCD phases, with possible connection
with $\alpha$-ZrCl$_3$ magnetism. 

\subsection*{Acknowledgments}
We thank Rafael Fernandes and Johannes Knolle for interesting
discussions and critical reading of our manuscript. W.M.H.N. acknowledges
the Royal Society for supporting this work through a Newton International
Fellowship. This work was supported by the Brazilian agency CNPq (E.C.A.). E.C.A acknowledges the hospitality of the International Institute
of Physics (IIP-UFRN), where part of this work was developed. Research at IIP-UFRN is supported by the Brazilian ministries MEC and MCTIC.

\bibliographystyle{apsrev4-1}
\bibliography{TBGv3}

\end{document}